# Computing Nash Equilibria: Approximation and Smoothed Complexity

Xi Chen [*]    Xiaotie Deng [†]    Shang-Hua Teng [‡]

**Abstract**

By proving that the problem of computing a $1/n^{\Theta(1)}$-approximate Nash equilibrium remains **PPAD**-complete, we show that the BIMATRIX game is not likely to have a fully polynomial-time approximation scheme. In other words, no algorithm with time polynomial in $n$ and $1/\epsilon$ can compute an $\epsilon$-approximate Nash equilibrium of an $n \times n$ bimatrix game, unless **PPAD** $\subseteq$ **P**. Instrumental to our proof, we introduce a new discrete fixed-point problem on a high-dimensional cube with a constant side-length, such as on an $n$-dimensional cube with side-length 7, and show that they are **PPAD**-complete. Furthermore, we prove that it is unlikely, unless **PPAD** $\subseteq$ **RP**, that the smoothed complexity of the Lemke-Howson algorithm or any algorithm for computing a Nash equilibrium of a bimatrix game is polynomial in $n$ and $1/\sigma$ under perturbations with magnitude $\sigma$. Our result answers a major open question in the smoothed analysis of algorithms and the approximation of Nash equilibria.

[*]Department of Computer Science, Tsinghua University, Beijing, P.R.China. xichen00@mails.tsinghua.edu.cn. Supported by National Natural Science Foundation of China (60135010, 60321002) and Chinese National Key Foundation Research and Development Plan (2004CB318108).

[†]Department of Computer Science, City University of Hong Kong, Hong Kong, P.R.China. deng@cs.cityu.edu.hk.

[‡]Department of Computer Science, Boston University, Boston, and Akamai Technologies Inc. Cambridge, USA. steng@cs.bu.edu. Partially supported by the NSF grants CCR-0311430 and ITR CCR-0325630. Part of the work was done while the author was visiting the City University of Hong Kong, Tsinghua University, and Microsoft Beijing Research Lab.

# 1  Introduction

The *two-person game* or the *bimatrix game* [13, 14] is perhaps the simplest form of non-cooperative games [19]. It is specified by two $m \times n$ matrices $\mathbf{A} = (a_{i,j})$ and $\mathbf{B} = (b_{i,j})$, where the $m$ rows represent the pure strategies for the first player, and the $n$ columns represent the pure strategies for the second player. In other words, if the first player chooses strategy $i$ and the second player chooses strategy $j$, then the payoffs to the first and the second players are $a_{i,j}$ and $b_{i,j}$, respectively.

Nash's theorem [19, 18] on non-cooperative games when specialized to bimatrix games states that there exists a profile of possibly mixed strategies so that neither player can gain by changing his/her (mixed) strategy, while the other player maintains her mixed strategies. Such a profile of strategies is called a *Nash equilibrium*.

Mathematically, a mixed strategy for the first player can be expressed by a column *probability vector* $\mathbf{x} \in \mathbb{R}^m$, that is, a vector with non-negative entries that sum to 1, while a mixed strategy for the second player is a probability vector $\mathbf{y} \in \mathbb{R}^n$. A Nash equilibrium is then a pair of probability vectors $(\mathbf{x}^*, \mathbf{y}^*)$ such that for all pairs of probability vectors $\mathbf{x} \in \mathbb{R}^m$ and $\mathbf{y} \in \mathbb{R}^n$,

$$(\mathbf{x}^*)^T \mathbf{A} \mathbf{y}^* \geq \mathbf{x}^T \mathbf{A} \mathbf{y}^* \quad \text{and} \quad (\mathbf{x}^*)^T \mathbf{B} \mathbf{y}^* \geq (\mathbf{x}^*)^T \mathbf{B} \mathbf{y}.$$

Note that the Nash equilibria of a bimatrix game $(\mathbf{A}, \mathbf{B})$ are invariant under positive scalings, that is, the bimatrix game $(c_1 \mathbf{A}, c_2 \mathbf{B})$ has the same set of Nash equilibria as the bimatrix game $(\mathbf{A}, \mathbf{B})$, as long as $c_1, c_2 > 0$. In addition, Nash equilibria are invariant under shifting, that is, for any constants $c_1$ and $c_2$, the bimatrix game $(c_1 + \mathbf{A}, c_2 + \mathbf{B})$ has the same set of Nash equilibria as the bimatrix game $(\mathbf{A}, \mathbf{B})$. Thus, we often normalize[1] the matrices $\mathbf{A}$ and $\mathbf{B}$ so that all their entries are between 0 and 1, or between -1 and 1.

The zero-sum two-person game [17] is a special case of the bimatrix game that satisfies $\mathbf{B} = -\mathbf{A}$. Moreover, an instance of the zero-sum game can be formulated as a linear program and hence can be solved in (weakly) polynomial time using, for example, the interior-point method [10]. However, the computational problem of the general two-person games and the existence proof of a Nash equilibrium have required different approaches from their zero-sum specializations.

One of the most popular methods to find a Nash equilibrium of a bimatrix game is the classic Lemke-Howson algorithm [14]. But recently, Savani and von Stengel [21] show that it needs an exponential number of steps in the worst case for the Lemke-Howson algorithm to reach an equilibrium solution. As the Lemke-Howson algorithm is somewhat connected with the simplex method for linear programming, this exponential worst-case lower bound did not completely damage the hope for an efficient solution of bimatrix games. After all, the exponential worst-case lower bound against almost all known simplex algorithms did not eliminate the possibility that linear programming can be solved in polynomial time [12, 10].

The following three optimistic conjectures summarize of our hope for the possible existence of an efficient method for computing an exact or an approximate Nash equilibrium of a bimatrix game.

---
[1] Normalization is needed in the definition of approximate Nash equilibrium to be discussed below.

1. **Polynomial 2-NASH Conjecture**: *there exists a (weakly) polynomial-time algorithm for computing a Nash equilibrium of a bimatrix game.*

2. **Fully Polynomial Approximate 2-NASH Conjecture**: One can relax the Nash condition and define an $\epsilon$-*approximate Nash equilibrium* to be a profile of mixed strategies such that no player can gain more than $\epsilon$ amount by changing his/her own strategy unilaterally. The conjecture[2] then states: *There exists an algorithm for computing an $\epsilon$-approximate Nash equilibrium in time polynomial in $n$, $m$, and $1/\epsilon$ for every $0 < \epsilon < 1$ and for every normalized game $(\mathbf{A}, \mathbf{B})$ with $\max_{i,j} \{|a_{i,j}|, |b_{i,j}|\} \leq 1$, where $a_{i,j}$ is the $(i,j)^{th}$ entry of $\mathbf{A}$ and $b_{i,j}$ is the $(i,j)^{th}$ entry of $\mathbf{B}$.*

3. **Smoothed 2-NASH Conjecture** [22]: *The problem of finding a Nash equilibrium of a bimatrix game is in smoothed polynomial time under $\sigma$-perturbations.*

   That is, for any pair of $m \times n$ matrices $\bar{\mathbf{A}} = (\bar{a}_{i,j})$ and $\bar{\mathbf{B}} = (\bar{b}_{i,j})$ with $|\bar{a}_{i,j}| \leq 1$ and $|\bar{b}_{i,j}| \leq 1$, let $\mathbf{A} = (a_{i,j})$ and $\mathbf{B} = (b_{i,j})$ be matrices where $a_{i,j} = \bar{a}_{i,j} + r^A_{i,j}$ and $b_{i,j} = \bar{b}_{i,j} + r^B_{i,j}$, with $r^A_{i,j}$ and $r^B_{i,j}$ being chosen independently and uniformly from $[-\sigma, \sigma]$. Then, a Nash equilibrium of the bimatrix game $(\mathbf{A}, \mathbf{B})$ can be solved in expected polynomial time in $m$, $n$, and $1/\sigma$.

   Ever since Spielman and Teng [23] established a polynomial upper bound on the smoothed complexity of the simplex algorithm with the shadow vertex pivoting rule, it has been frequently asked *whether the smoothed complexity of another classic algorithm, the Lemke-Howson algorithm for bimatrix games, is polynomial.*

In the final installment of a series of exciting developments initiated by Daskalakis, Goldberg and Papadimitriou [7], Chen and Deng [3] show that the first conjecture is unlikely to be true. More specifically, they prove that the problem of computing a Nash equilibrium is **PPAD**-complete for bimatrix games in which each player has $n$ pure strategies.

In this paper, we show that it is unlikely that the second and third conjectures above are true — we show that it remains **PPAD**-complete to compute a $1/n^{\Theta(1)}$-approximate Nash equilibrium of an $n \times n$ normalized bimatrix game. Consequently, we show that it is unlikely that there is an algorithm for computing an $\epsilon$-approximate Nash equilibrium with complexity polynomial in $n$ and $1/\epsilon$. Thus, it is unlikely that the $n^{O(\log n/\epsilon^2)}$-time $\epsilon$-approximate result of Lipton, Markakis, and Mehta [15] can be dramatically improved to poly($n, 1/\epsilon$).

By exploiting the connection between the complexity of approximate Nash equilibria and the smoothed complexity of Nash equilibria, we show that, unless **PPAD** $\subseteq$ **RP**, it is unlikely that the smoothed complexity of the classic Lemke-Howson algorithm or any algorithm for bimatrix games is polynomial in $n$ and $1/\sigma$ under perturbations with magnitude $\sigma$. Thus, the average-case polynomial time result of Barany, Vempala, and Vetta [2] is not likely extendible to the smoothed model.

---

[2] Because the notion of the $\epsilon$-approximate Nash equilibria is defined in the additive fashion, to study its complexity, it is important to consider bimatrix games with normalized matrices. That is, the absolute value of each entry in the matrices is bounded, for example by 1. Earlier work on this subject by Lipton, Markakis, and Mehta [15] used a similar normalization. Although, exact Nash equilibria of $(\mathbf{A}, \mathbf{B})$ are invariant under positive scalings, each $\epsilon$-approximate Nash equilibrium $(\mathbf{x}, \mathbf{y})$ of $(\mathbf{A}, \mathbf{B})$, becomes a $c \cdot \epsilon$-approximate Nash equilibrium of the bimatrix game $(c\mathbf{A}, c\mathbf{B})$ for $c > 0$. It is worthwhile noticing that $\epsilon$-approximate Nash equilibria are invariant under shifting, that is, for any constants $c_1$ and $c_2$, the bimatrix game $(c_1 + \mathbf{A}, c_2 + \mathbf{B})$ has the same set $\epsilon$-Nash equilibria as the bimatrix game $(\mathbf{A}, \mathbf{B})$. To properly define the perturbation magnitudes in the next conjecture, we also consider normalized bimatrix games.

Our approach depends on a new discrete version of the Brouwer's fixed point. Iimura introduced a concept of direction-preserving functions to derive a discrete fixed point theorem [9]. The concept was applied by Chen and Deng [4] to derive a matching upper and lower bound for finding an approximate fixed point for any fixed dimension higher than two. A notion of the "bad" cube was utilized in the analysis. Until this point, the fixed point for discrete functions exists only for restricted classes of discrete functions. Later, Daskalakis, Goldberg and Papadimitriou introduced a discrete concept of fixed points in term of the values of the function at the eight vertices of a three dimensional unit cube and proved that the problem for computing this type of 3D discrete Brouwer's fixed points is **PPAD**-complete [7]. This invention allowed a discrete fixed point to exist once a boundary condition is satisfied by the function without any further restriction on the function itself otherwise. Using an improvement on the SPERNER problem, Chen and Deng improved the result to the two dimensional case [3]. The new discrete version of the fixed-point problem has been helpful in proving the **PPAD**-completeness of 4-NASH[3] [7], and the extension to 3-NASH [5, 8], as well as the **PPAD**-complete proof of the exact Nash equilibrium of two-person games [6].

However, those previous hardness proofs apply only to the computation of an approximate Nash equilibrium within an exponentially small approximation parameter.

For example, the result of Chen and Deng states that finding a $2^{-\Theta(n)}$-approximate Nash equilibrium of a bimatrix game, in which each player has $n$ pure strategies, is **PPAD**-complete. Their proof does not apply to the computation of $1/n^{\Theta(1)}$-approximate Nash equilibria, because the approximation ratio decreases geometrically in the number of bits under consideration for the fixed point problem. In order to extend these hardness results to the computation of a $1/n^{\Theta(1)}$-approximate Nash equilibrium, we need a reduction from a **PPAD**-complete fixed point problem with a constant or $O(\log n)$ number of bits, which seems inconceivable in previous approaches.

As an instrumental step in achieving our result, we introduce a family of high-dimensional discrete fixed-point problems. The high-dimensional spaces provide an effecitve trade-off between the dimension and the side length of the hypergrid inside the cube that defines the search space. In particular, we consider a discrete fixed-point problem associated with the integer lattice points of the $(\underbrace{8 \times 8 \times ... \times 8}_{n})$ $n$-dimensional cube.

Fortunately and somewhat suprisingly, the discrete fixed-point problem is still **PPAD**-complete on this seemly small cube. A cube with a constant side-length in high dimension still contains an exponential number of integer lattice vertices. We show that as long as its side-length is reasonably large (e.g., is equal to eight) the cube has enough space and flexibility to fit a complex proof structure of a **PPAD**-complete problem.

---

[3]Note that entries of a Nash equilibrium of a four-person game may not be rational, even with rational payoff entries. So, in general, one can only hope to compute an approximate Nash equilibrium of a four-person games. Formally, the hardness result of Daskalakis, Goldberg, and Papadimitriou states: finding a $2^{-\Theta(n)}$-approximate Nash equilibrium of a normalized four-person game in which each players has $n$ strategies is **PPAD**-complete.

However, discrete fixed-point problems in high dimensions come with their own challenges. In particular, the exponential number of vertices in a unit cube create new difficulties to the efficiency of the reduction techniques. Our definition of the family of high-dimensional discrete fixed-point problems itself reflects a simple example of the necessity of handling that hurdle. We develop several new techniques that enable us to carry through the reduction. A particularly important one is a new averaging method to overcome the curse of high dimensionality in the closing loop of the reduction that connects the computation of Nash equilibria in a bimatrix game with the high-dimensional fixed-point problem.

As the fixed-points and Nash equilibria are fundamental to many other search and optimization problems, our results and techniques may have a broader scope of applications and implications. For example, our hardness results on bimatrix games can be naturally extended to $r$-person matrix games and $r$-graphical games for any fixed $r$.

## 1.1 Notations

We will use bold lower-case Roman letters such as $\mathbf{x}$, $\mathbf{a}$, $\mathbf{b}_j$ to denote vectors. Whenever a vector, say $\mathbf{a} \in \mathbb{R}^n$ is present, its components will be denoted by lower-case Roman letters with subscripts, such as $a_1, \ldots, a_n$. Matrices are denoted by bold upper-case Roman letters such as $\mathbf{A}$ and scalars are usually denoted by lower-case roman letters, but sometime by upper-case Roman letters such as $M$, $N$, and $K$. The $(i,j)^{th}$ entry of a matrix $\mathbf{A}$ is denoted by $a_{i,j}$. Depending on the context, we may use $\mathbf{a}_i$ to denote the $i^{th}$ row or the $i^{th}$ column of $\mathbf{A}$.

We now enumerate some other notations that are used in this paper.

- $\mathbb{Z}_+^d$: the set of $d$-dimensional vectors with positive integer entries;

- $\mathbb{Z}_{[a,b]}^d$: the set $\left\{\mathbf{p} \in \mathbb{Z}^d \mid a \leq p_i \leq b, \forall 1 \leq i \leq d\right\}$, for integers $a \leq b$.

- $\mathbb{R}_{[a:b]}^{m \times n}$: the set of all $m \times n$ matrices with real entries between $a$ and $b$. For example, $\mathbb{R}_{[-1:1]}^{m \times n}$ is the set of the $m \times n$ matrices with entries whose absolute values are at most 1.

- $\mathbb{P}^n$: the set of all probability vectors in $n$ dimensions. For example, $\mathbf{x} \in \mathbb{P}^n$ means $\sum_{i=1}^n x_i = 1$ and $x_i \geq 0$ for all $1 \leq i \leq n$.

- $\langle \mathbf{a} | \mathbf{b} \rangle$: the dot-product of two vectors in the same dimension.

- $\mathbf{e}_i$: the unit vector whose $i^{th}$ entry is equal to 1 and all other entries are zeros.

## 1.2 Organization of the Paper

In Section 2, we review some complexity classes that will be used in this paper. We also introduce a family of high-dimensional discrete fixed-point problems. In Section 3, we prove that this family of high-dimensional discrete-fixed point problems is **PPAD**-complete, independent of the dimension of the search space. In Section 4, we show that the problem of computing a $1/n^{\Theta(1)}$-approximate Nash equilibrium is **PPAD**-complete. In Section 5, we consider the smoothed complexity of the bimatrix game. Finally, in Section 6, we conclude the paper with some open questions inspired by our study.

## 2 PPAD and High-Dimensional Brouwer's Fixed Points

In this section, we review the complexity class **PPAD** introduced by Papadimitriou [20] and define a family of search problems about the computation of a Brouwer's fixed point in a high dimensional space. In particular, we allow the value of the dimension to be included in the input size. The consideration of high-dimensional fixed point problem is instrumental to establish our main result.

### 2.1 TFNP and PPAD

Let $R \subset \{0,1\}^* \times \{0,1\}^*$ be a binary relation over $\{0,1\}^*$. $R$ is *polynomially balanced* if there exists a polynomial $p$ such that for all pairs $(x,y) \in R$, $|y| \leq p(|x|)$. In addition, $R$ is a *polynomial-time relation* if for each pair $(x,y)$, one can decide whether or not $(x,y) \in R$ in time polynomial in $|x| + |y|$.

One can define the **NP** search problem $Q_R$ specified by $R$ as: Given an input string $x \in \{0,1\}^*$, decide whether there exists a $y \in \{0,1\}^*$ such that $(x,y) \in R$, and if such $y$ exists, return $y$, otherwise, return a special string called "no".

A relation $R$ is *total* if for every string $x \in \{0,1\}^*$, there exists a $y$ such that $(x,y) \in R$. Following Megiddo and Papadimitriou [16], **TFNP** denotes the class of all **NP** search problems defined by total relations.

**Definition 1 (Polynomial Reduction).** *A search problem $Q_{R_1} \in$ **TFNP** is polynomial-time reducible to another $Q_{R_2} \in$ **TFNP** if there exists a pair of polynomial-time computable functions $(f,g)$ such that, for every input $x$ of $R_1$, if $y$ satisfies $(f(x),y) \in R_2$, then $(x,g(y)) \in R_1$. $Q_{R_1}$ and $Q_{R_2}$ are then polynomial-time equivalent (or simply, equivalent) if $Q_{R_2}$ is also reducible to $Q_{R_1}$.*

An important and interesting sub-class of **TFNP** is the complexity class **PPAD** [20]. It is the set of total search problems that are polynomial-time reducible to the following search problem, LEAFD.

**Definition 2 (LEAFD).** *The input instance of LEAFD is a pair $(M, 0^n)$ where $M$ is the description of a polynomial-time Turing machine satisfying the following two conditions:*

1. *for every $v \in \{0,1\}^n$, $M(v)$ is an ordered pair $(u_1, u_2)$ where $u_1, u_2 \in \{0,1\}^n \cup \{$ "no"$\}$.*

2. *$M(0^n) = ($ "no"$, 1^n)$ and the first component of $M(1^n)$ is $0^n$.*

*This instance defines a directed graph $G = (V, E)$ where $V = \{0,1\}^n$ and a pair $(u,v) \in E$ if and only if $v$ is the second component of $M(u)$ and $u$ is the first component of $M(v)$.*

*The output of this problem is a directed leaf of $G$ other than $0^n$, where a vertex of $V$ is a directed leaf if its out-degree plus in-degree equals one.*

In other words, a **PPAD** instance is defined by a directed graph (of exponential size) $G = (V, E)$, where both the in-degree and the out-degree of every node are at most 1, together with a pair of polynomial-time functions, $P : V \to V \cap \{$ "no"$\}$ and $S : V \to V \cap \{$ "no"$\}$ that compute the predecessor and successor of each vertex, respectively. A pair $(u,v)$ appears in $E$ only if $P(v) = u$ and $S(u) = v$. In addition, a starting source vertex $0^n$ with in-degree 0 and out-degree 1 is given. The required output is another vertex with in-degree 1 and out-degree 0 (a *sink*) or with in-degree 0 and out-degree 1 (another *source*).

Many important problems, such as the search versions of Brouwer's fixed-point theorem, Kakutani's fixed-point theorem, Smith's theorem and Borsuk-Ulam theorem, have been identified to be in the class **PPAD** [20].

The totality of **PPAD** is guaranteed by the following fact: in a directed graph where the in-degree and out-degree of every vertex are at most one, if there exists a source, there must be a sink or another source[4]. Simply from its definition, LEAFD is complete for **PPAD**.

## 2.2  Problem BROUWER$^f$

One of the most important **PPAD** problems concerns the task of search for a discrete Brouwer's fixed point. In this subsection, we define this class of search problems in various dimensions. A discrete version of the Brouwer's fixed point theorem provides key to obtain our main results. In order to handle the exponential exploration of vertices in high dimensional cube, we need to define an efficient discrete version. We will introduce some notations — some elementary and some geometric — to facilitate our discussion.

To define our high dimensional Brouwer's fixed point problems, we need a notion of *well-behaved* functions (please note that this is not the function for the fixed point problem) to parameterize the shape of the search space. We say an integer function $f(n)$ is *well-behaved* if it is polynomial-time computable and there exists a constant $n_0$ such that $3 \leq f(n) \leq n/2$ for every integer $n \geq n_0$.

For example, $f_1(n) = 3$, $f_2(n) = \lfloor n/2 \rfloor$, $f_3(n) = \lfloor n/3 \rfloor$ and $f_4(n) = \lfloor \log n \rfloor$ are all well-behaved functions.

For each $\mathbf{p} \in \mathbb{Z}^d$, let $K_{\mathbf{p}}$ denote the following unit hypercube incident to $\mathbf{p}$, that is,

$$K_{\mathbf{p}} = \left\{ \mathbf{q} \in \mathbb{Z}^d \ \middle| \ q_i = p_i \text{ or } p_i + 1, \ \forall \ 1 \leq i \leq d. \right\}.$$

For a positive integer $d$ and a vector $\mathbf{r} \in \mathbb{Z}_+^d$, let

$$A_{\mathbf{r}}^d = \left\{ \mathbf{p} \in \mathbb{Z}^d \ \middle| \ 0 \leq p_i \leq r_i - 1, \forall \ 1 \leq i \leq d. \right\}$$

be the *hyper-grid* with side length given by $\mathbf{r}$. The *boundary* of $A_{\mathbf{r}}^d$, $\partial \left( A_{\mathbf{r}}^d \right)$, is then the set of integer lattice points $\mathbf{p} \in A_{\mathbf{r}}^d$ with $p_i \in \{0, r_i - 1\}$ for some $i : 1 \leq i \leq d$.

For each $\mathbf{r} \in \mathbb{Z}_+^d$, let $\text{Size}[\mathbf{r}] = \sum_{1 \leq i \leq d} \lceil \log(r_i + 1) \rceil$ denote the number of bits necessary and sufficient to represent $\mathbf{r}$.

---
[4]**PPAD** is the directed version of class **PPA**, also introduced by Papadimitriou [20]. **PPA** is a class of search problems defined based on the Polynomial Parity Argument: Given a finite graph consisting of lines and cycles, there must be an even number of vertices with degree 1. A **PPA** instance specifies an exponentially-sized undirected graph in which each vertex has degree no more than 2, a polynomial-time computable neighboring function, and a vertex, say $0^n$, with degree 1. The goal of its search problem is to find another degree 1 vertex.

**Definition 3 (Brouwer-Mapping Circuit).** *For a positive integer $d$ and $\mathbf{r} \in \mathbb{Z}_+^d$, a Boolean circuit $C$ is a Brouwer-mapping circuit with parameters $d$ and $\mathbf{r}$ if it has exactly $\text{Size}\,[\mathbf{r}]$ input bits and $2d$ output bits $\Delta_1^+, \Delta_1^- \ldots \Delta_d^+, \Delta_d^-$.*

*Moreover, $C$ is a valid Brouwer-mapping circuit if*

- *for every $\mathbf{p} \in A_{\mathbf{r}}^d$, the set of $2d$ output bits evaluated at $\mathbf{p}$ falls into one of the following cases:*
  - *case 1: $\Delta_1^+ = 1$ and all other bits are 0;*
  - *...*
  - *case $d$: $\Delta_d^+ = 1$ and all other bits are 0;*
  - *case $d+1$: for every $i : 1 \leq i \leq d$, $\Delta_i^+ = 0$ and $\Delta_i^- = 1$.*

- *for every $\mathbf{p} \in \partial\left(A_{\mathbf{r}}^d\right)$, if there exists an $i : 1 \leq i \leq d$ such that $p_i = 0$, letting $i_{\max} = \max\{i \mid p_i = 0\}$, then the output bits satisfy the $i_{\max}^{th}$ case, otherwise, (when none of the $p_i$ is 0 and some are $r_i - 1$), the output bits satisfy the $d+1^{st}$ case.*

**Definition 4 (Brouwer Color Assignment and Panchromatic Simplex).** *Suppose $C$ is a valid Brouwer-mapping circuit with parameter $d$ and $\mathbf{r}$. Circuit $C$ defines a color assignment: $\text{Color}_C : A_{\mathbf{r}}^d \to \{1, 2, \ldots d, \text{"red"}\}$, where "red" is a special color, according to the following rule: For each point $\mathbf{p} \in A_{\mathbf{r}}^d$, if the set of output bits of $C$ evaluated at $\mathbf{p}$ satisfies the $i^{th}$ case where $1 \leq i \leq d$, then $\text{Color}_C\,[\mathbf{p}] = i$; otherwise, the output bits of $C$ satisfy the $d+1^{th}$ case and $\text{Color}_C\,[\mathbf{p}] = \text{"red"}$.*

*A subset $P \subset A_{\mathbf{r}}^d$ is accommodated if there exists a point $\mathbf{p} \in A_{\mathbf{r}}^d$ such that $P \subset K_{\mathbf{p}}$. A set $P \subset A_{\mathbf{r}}^d$ is a panchromatic set or a panchromatic simplex of $C$ if it is an accommodated set of $d+1$ points assigned with all $d+1$ colors.*

For each well-behaved function $f$, we define a discrete Brouwer fixed point problem as following.

**Definition 5 (BROUWER$^f$).** *For a well-behaved function $f$ and a parameter $n$, let $m = f(n)$ and $d = \lceil n/f(n) \rceil$. An input instance of BROUWER$^f$ is a pair $(C, 0^n)$ where $C$ is a valid Brouwer-mapping circuit[5] with parameter $d$ and $\mathbf{r}$ where $r_i = 2^m, \forall\, i : 1 \leq i \leq d$. The output of this search problem is then a panchromatic simplex of $C$.*

The following property will be useful for the next section.

---

[5]Formally, a polynomially balanced and polynomial-time relation $R$ is necessary in order to define a search problem in **TFNP**. As the input is an arbitrary string $\mathbf{x}$, it may happen that its $C$ component is not a description of a valid Brouwer-mapping circuit. It is even possible that $\mathbf{x}$ cannot be decoded into a boolean circuit following by a sequence of 0's. We can embed valid Brouwer-mapping circuits into a polynomially balanced and polynomial-time computable relation as following: In advance, we choose a "standard" valid Brouwer-mapping circuit $C^*$ with parameters $d$ and $\mathbf{r}$ and size polynomial in $n$. If a string $\mathbf{x}$ of $\{0,1\}^*$ can not be decoded into a Brouwer-mapping circuit, followed by a sequence of 0's, then we include $(\mathbf{x}, 0^1)$ in $R$. Otherwise, assume $\mathbf{x}$ is decoded into $(C, 0^n)$. We couple $C$ with $C^*$, in case $C$ is not a valid Brouwer-mapping circuit, to define a possibly new circuit $C'$. If the output bits of $C$ evaluated at $\mathbf{p} \in A_{\mathbf{r}}^d$ is invalid, then $C'$ invokes $C^*$ and output according to $C^*$. We include $(\mathbf{x}, \mathbf{y})$ in $R$ if $\mathbf{y}$ can be decoded into a panchromatic simplex of $C'$. One can verify that $R$ is indeed polynomially balanced and polynomial-time computable. In addition, $C' = C$ for all valid $C$.

**Property 1 (Boundary Continuity).** *Let $C$ be a valid Brouwer-mapping circuit with parameters $d$ and $\mathbf{r}$, and $\mathbf{p}, \mathbf{p}' \in \partial\left(A_\mathbf{r}^d\right)$. If $\mathbf{p}' = \mathbf{p} + \mathbf{e}_t$ for some $1 \le t \le d$ and $1 \le p_t \le r_t - 2$, then $\mathrm{Color}_C[\mathbf{p}] = \mathrm{Color}_C[\mathbf{p}']$.*

*Proof.* By the definition, if $C$ is a valid Brouwer-mapping circuit $C$ with parameters $d$ and $\mathbf{r}$, then for each $\mathbf{p} \in \partial\left(A_\mathbf{r}^d\right)$, $\mathrm{Color}_C[\mathbf{p}]$ satisfies

- if there exists an $i : 1 \le i \le d$ such that $p_i = 0$, then $\mathrm{Color}_C[\mathbf{p}] = \max\{i \mid p_i = 0\}$; otherwise, when none of $p_i$ is $0$ and some of them are $r_i - 1$, $\mathrm{Color}_C[\mathbf{p}] = $ "red".

Thus, if $1 \le p_t \le r_t - 2$, then $\mathrm{Color}_C[\mathbf{p}] = \mathrm{Color}_C[\mathbf{p}']$. □

**Lemma 1.** *For any well-behaved function $f$, search problem $\mathrm{BROUWER}^f$ is in $\mathbf{PPAD}$.*

*Proof.* See appendix. □

## 3 PPAD-Completeness of BROUWER$^f$

In this section, we show that for any well-behaved function $f$, BROUWER$^f$ is **PPAD**-complete. The hardness result essentially states that the complexity of finding a Brouwer's fixed point is independent of the shape or dimension of the domain. Instead, it is dominated by the number of points and hence the number of unit hypercubes in the domain.

Recall $f_1(n) = 3$, $f_2(n) = \lfloor n/2 \rfloor$, $f_3(n) = \lfloor n/3 \rfloor$ are all well-behaved functions. It was recently shown that both BROUWER$^{f_2}$ [3] and BROUWER$^{f_3}$ [7] are **PPAD**-complete. Our reduction is from the two-dimensional BROUWER$^{f_2}$. In section 4, we reduce BROUWER$^{f_1}$, where $f_1(n) = 3$, to the computation of a $1/n^{\Theta(n)}$-approximate Nash equilibrium of bimatrix games.

### 3.1 Reductions among Coloring Triples

The basic idea of our reduction is to iteratively embed an instance of BROUWER$^{f_2}$ into higher dimension space. We will use the following concept to describe such embedding processes.

**Definition 6 (Coloring Triples).** *A triple $T = (C, d, \mathbf{r})$ is said to be a coloring triple if $\mathbf{r} \in \mathbb{Z}^d$ with $r_i \ge 7$ for all $i : 1 \le i \le d$ and $C$ is a valid Brouwer-mapping circuit with parameters $d$ and $\mathbf{r}$.*

We start with three lemmas that are important for reduction. In each lemma, we introduce a transformation that embeds a given coloring triple $T$ into a somewhat larger coloring triple $T'$ such that from a panchromatic simplex of $T'$, we can compute a panchromatic simplex of $T$ efficiently.

We will use $\mathrm{Size}[C]$ to denote the number of gates plus the number input and output variables in a Boolean circuit $C$.

**Lemma 2 ($\mathbf{L}^1(T, t, u)$: Padding a Dimension).** *Given a coloring triple $T = (C, d, \mathbf{r})$ and integers $1 \le t \le d$ and $u > r_t$, we can construct a new coloring triple $T' = (C', d, \mathbf{r}')$ that satisfies the following two conditions.*

| **Color$_{C'}$[p] of a point p $\in A^d_{\mathbf{r'}}$ assigned by $\mathbf{L}^1(T,t,u)$** |
|---|
| 1: **if** $\mathbf{p} \in \partial \left( A^d_{\mathbf{r'}} \right)$ **then** |
| 2:    **if** there exists $i$ such that $p_i = 0$ **then** |
| 3:       Color$_{C'}$[p] $= i_{\max} = \max\{\, i \mid p_i = 0 \,\}$ |
| 4:    **else** |
| 5:       Color$_{C'}$[p] = red |
| 6: **else if** $p_t \leq r_t$ **then** |
| 7:    Color$_{C'}$[p] = Color$_C$[p] |
| 8: **else** |
| 9:    Color$_{C'}$[p] = red |

Figure 1: How $\mathbf{L}^1(T,t,u)$ extends the color assignment

**A.** *For all $i : 1 \leq i \neq t \leq d$, $r'_i = r_i$, and $r'_t = u$. In addition, there exists a polynomial $g_1(n)$ such that $Size[C'] = Size[C] + O(g_1(Size[\mathbf{r'}]))$ and $T'$ can be computed in time polynomial in $Size[C']$. We write $T' = \mathbf{L}^1(T,t,u)$.*

**B.** *From each panchromatic simplex $P'$ of coloring triple $T'$, we can compute a panchromatic simplex $P$ of $T$ in polynomial time.*

*Proof.* We define circuit $C'$ by its color assignment in Figure 1. Property **A** is true according to this definition.

To show Property **B**, let $P' \subset K_{\mathbf{p}}$ be a panchromatic simplex of $T'$. We first note that $p_t \leq r_t - 1$, because had $p_t > r_t - 1$, $K_{\mathbf{p}}$ would not contain color $t$ according to the color assignment, Thus, it follows from Color$_{C'}$[q] = Color$_C$[q] for each $\mathbf{q} \in A^d_{\mathbf{r}}$ that $P'$ is also be a panchromatic simplex of the coloring triple $T$. □

**Lemma 3 ($\mathbf{L}^2(T,u)$: Adding a Dimension).** *Given a coloring triple $T = (C, d, \mathbf{r})$ and an integer $u \geq 7$, we can construct a new coloring triple $T' = (C', d+1, \mathbf{r'})$ that satisfies the following conditions.*

**A.** *For all $i : 1 \leq i \leq d$, $r'_i = r_i$, and $r'_{d+1} = u$. Moreover, there exists a polynomial $g_2(n)$ such that $Size[C'] = Size[C] + O(g_2(Size[\mathbf{r'}]))$ and $T'$ can be computed in time polynomial in $Size[C']$. We write $T' = \mathbf{L}^2(T,u)$.*

**B.** *From each panchromatic simplex $P'$ of coloring triple $T'$, we can compute a panchromatic simplex $P$ of $T$ in polynomial time.*

*Proof.* For each point $\mathbf{p} \in A^{d+1}_{\mathbf{r'}}$, we use $\bar{\mathbf{p}}$ to denote the point $\mathbf{q} \in A^d_{\mathbf{r}}$ with $q_i = p_i, \forall i : 1 \leq i \leq d$. The color assignment of circuit $C'$ is given in Figure 2. Clearly, Property **A** is true.

To prove property **B**, let $P' \subset K_{\mathbf{p}}$ be a panchromatic simplex of $T'$. We note that $p_{d+1} = 0$, for otherwise, $K_{\mathbf{p}}$ does not contain color $d+1$. Note also Color$_{C'}$[q'] $= d+1$ for every $\mathbf{q} \in A^{d+1}_{\mathbf{r'}}$ with $q_{d+1} = 0$. Thus, for every $\mathbf{q} \in P'$ with Color$_{C'}$[q] $\neq d+1$, we

---
**Color$_{C'}$ [p] of a point p $\in A_{\mathbf{r}'}^{d+1}$ assigned $\mathbf{L}^2(T,u)$**
---
1: **if** $\mathbf{p} \in \partial\left(A_{\mathbf{r}'}^{d}\right)$ **then**
2:     **if** there exists $i$ such that $p_i = 0$ **then**
3:         Color$_{C'}$ [p] $= i_{\max} = \max\{\,i \mid p_i = 0\,\}$
4:     **else**
5:         Color$_{C'}$ [p] $=$ red
6: **else if** $p_{d+1} = 1$ **then**
7:     Color$_{C'}$ [p] $=$ Color$_C$ [$\bar{\mathbf{p}}$]
8: **else**
9:     Color$_{C'}$ [p] $=$ red
---

Figure 2: How $\mathbf{L}^2(T, u)$ extends the color assignment

have $q_{d+1} = 1$. So, because Color$_{C'}$ [q] $=$ Color$_C$ [$\bar{\mathbf{q}}$] for each $\mathbf{q} \in A_{\mathbf{r}'}^{d+1}$ with $q_{d+1} = 1$, $P = \{\bar{\mathbf{q}} \mid \mathbf{q} \in P'$ and Color$_{C'}$ [p] $\neq d+1\}$ is a panchromatic simplex of coloring triple $T$. $\square$

**Lemma 4 ($\mathbf{L}^3(T, t, a, b)$: Snake Embedding).** *Given a coloring triple $T = (C, d, \mathbf{r})$ and integer $1 \leq t \leq d$, if $r_t = a(2b+1) + 5$ for some integers $a, b \geq 1$, then we can construct a new triple $T' = (C', d+1, \mathbf{r}')$ that satisfies the following conditions.*

**A.** *For $i : 1 \leq i \neq t \leq d$, $r_i' = r_i$ and $r_t' = a + 5$ and $r_{d+1}' = 4b + 3$. Moreover, there exists a polynomial $g_3(n)$ such that $Size[C'] = Size[C] + O(g_3(Size[\mathbf{r}']))$ and $T'$ can be computed in time polynomial in $Size[C']$. We write $T' = \mathbf{L}^3(T, t, a, b)$.*

**B.** *From each panchromatic simplex $P'$ of coloring triple $T'$, we can computer a panchromatic simplex $P$ of $T$ in polynomial time.*

*Proof.* Consider the domain of $A_{\mathbf{r}}^{d} \subset \mathbb{Z}^d$ and $A_{\mathbf{r}'}^{d+1} \subset \mathbb{Z}^{d+1}$ of our coloring triples. We form the reduction $\mathbf{L}^3(T, t, a, b)$ in three steps. First, we define a $d$-dimensional set $W \subset A_{\mathbf{r}'}^{d+1}$ that is large enough to contain $A_{\mathbf{r}}^d$. Second, we define a map $\psi$ that (implicitly) specifies an embedding of $A_{\mathbf{r}}^d$ into $W$. Finally, we build a circuit $C'$ for $A_{\mathbf{r}'}^{d+1}$ and show that from each panchromatic simplex of $C'$, we can in polynomial time compute a pancrhomatic simplex of $C$.

A two dimensional view of $W$ is illustrated in Figure 3. We use a snake-pattern to realizes the longer $t^{th}$ dimension of $A_{\mathbf{r}}^d$ in the two-dimensional space defined by the shorter $t^{th}$ and $d+1^{st}$ dimensions of $A_{\mathbf{r}'}^{d+1}$. Formally, $W$ consists the set of points $\mathbf{p} \in A_{\mathbf{r}'}^{d+1}$ satisfying $1 \leq p_{d+1} \leq 4b+1$ and

    if $p_{d+1} = 1$, then $2 \leq p_t \leq a + 4$ and if $p_{d+1} = 4b + 1$, then $0 \leq p_t \leq a + 2$;

    if $p_{d+1} = 4(b - i) - 1$ where $0 \leq i \leq b - 1$, then $2 \leq p_t \leq a + 2$;

    if $p_{d+1} = 4(b - i) - 3$ where $0 \leq i \leq b - 2$, then $2 \leq p_t \leq a + 2$;

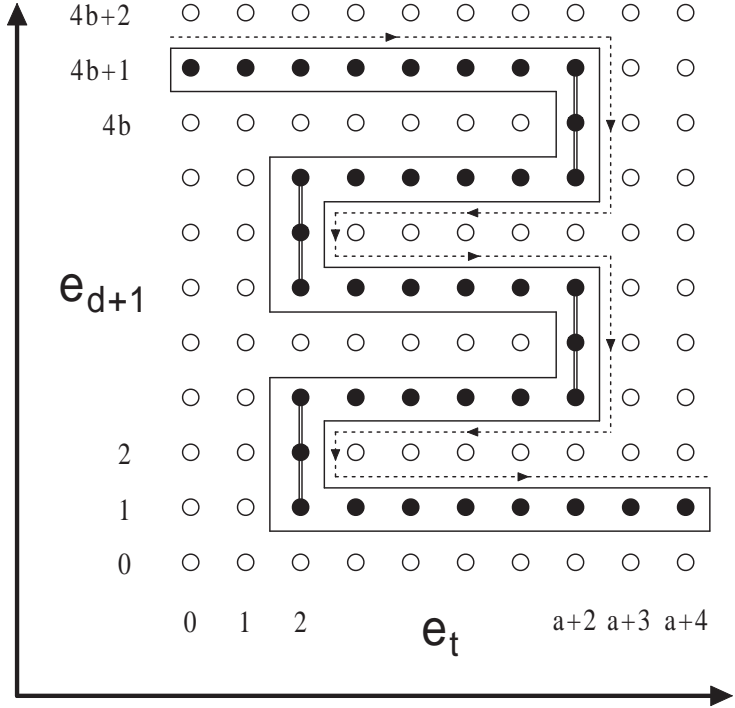

Figure 3: The two dimensional view of set $W \subset A_{\mathbf{r}'}^{d+1}$

if $p_{d+1} = 4(b-i) - 2$ where $0 \leq i \leq b-1$, then $p_t = 2$;

if $p_{d+1} = 4(b-i)$ where $0 \leq i \leq b-1$, then $p_t = a+2$.

To build $T'$, we embed the coloring triple $T$ into $W$. This embedding is implicitly given by a natural surjective map $\psi$ from $W$ to $A_{\mathbf{r}}^d$, a map that will plays a vital role in our construction and analysis. For each $\mathbf{p} \in W$, we use $\mathbf{p}[m]$ to denote the point $\mathbf{q}$ in $\mathbb{Z}^d$ such that $q_t = m$ and $q_i = p_i$, $\forall i : 1 \leq i \neq t \leq d$. We define $\psi(\mathbf{p})$ according to the following cases:

if $p_{d+1} = 1$, then $\psi(\mathbf{p}) = \mathbf{p}[2ab + p_t]$ and if $p_{d+1} = 4b+1$, then $\psi(\mathbf{p}) = \mathbf{p}[p_t]$;

if $p_{d+1} = 4(b-i) - 1$ where $0 \leq i \leq b-1$, then $\psi(\mathbf{p}) = \mathbf{p}[(2i+2)a + 4 - p_t]$;

if $p_{d+1} = 4(b-i) - 3$ where $0 \leq i \leq b-2$, then $\psi(\mathbf{p}) = \mathbf{p}[(2i+2)a + p_t]$;

if $p_{d+1} = 4(b-i) - 2$ where $0 \leq i \leq b-1$, then $\psi(\mathbf{p}) = \mathbf{p}[(2i+2)a + 2]$;

if $p_{d+1} = 4(b-i)$ where $0 \leq i \leq b-1$, then $\psi(\mathbf{p}) = \mathbf{p}[(2i+1)a + 2]$.

Essentially, we map $W$ bijectively to $A_{\mathbf{r}}^d$ along its $t^{th}$ dimension with exception that when the snake pattern of $W$ is making a turn, we stop the advance in $A_{\mathbf{r}}^d$, and continue the advance after it completes the turn. Let $\psi_i(\mathbf{p})$ denote the $i^{th}$ component of $\psi(\mathbf{p})$. Our embedding scheme guarantees the following important property of $\psi$.

**Color$_{C'}$ [p] of a point $\mathbf{p} \in A_{\mathbf{r}'}^{d+1}$ assigned $\mathbf{L}^3(T,t,a,b)$**

1: **if** $\mathbf{p} \in W$ **then**
2:     Color$_{C'}$ [p] = Color$_C$ [$\psi(\mathbf{p})$]
3: **else if** $\mathbf{p} \in \partial\left(A_{\mathbf{r}'}^{d+1}\right)$ **then**
4:     **if** there exists $i$ such that $p_i = 0$ **then**
5:         Color$_{C'}$ [p] = $i_{\max} = \max\{\, i \mid p_i = 0 \,\}$
6:     **else**
7:         Color$_{C'}$ [p] = red
8: **else if** $p_{d+1} = 4i$ where $1 \leq i \leq b$ and $1 \leq p_t \leq a+1$ **then**
9:     Color$_{C'}$ [p] = $d+1$
10: **else if** $p_{d+1} = 4i+1, 4i+2$ or $4i+3$ where $0 \leq i \leq b-1$ and $p_t = 1$ **then**
11:     Color$_{C'}$ [p] = $d+1$
12: **else**
13:     Color$_{C'}$ [p] = red

Figure 4: The color assignment of $\mathbf{L}^3(T,t,a,b)$

**Property 2 (Boundary Preserving).** *Let $\mathbf{p}$ be a point in $W \cap \partial\left(A_{\mathbf{r}'}^{d+1}\right)$. If there exists $i$ such that $p_i = 0$, then $\max\{\, i \mid p_i = 0 \,\} = \max\{\, i \mid \psi_i(\mathbf{p}) = 0 \,\}$. Otherwise, all entries of $\mathbf{p}$ are non-zero and there exists $l$ such that $p_l = r_l' - 1$, in which case, all entries of $\psi_i(\mathbf{p})$ are nonzero and $\psi_l(\mathbf{p}) = r_l - 1$.*

Circuit $C'$ specifies a color assignment of $A_{\mathbf{r}'}^{d+1}$ according to Figure 4. This circuit is derived from circuit $C$ and map $\psi$. By Property 2, we can verify that $C'$ is a valid Brouwer-mapping circuit with parameters $d+1$ and $\mathbf{r}'$.

Property **A** of the lemma follows directly from our construction. In order to establish Property **B** of the lemma, we prove the following collection of statements to cover all possible cases of the given panchromatic simplex of $T'$. In the following statements, $P'$ is a panchromatic simplex of $T'$ in $A_{\mathbf{r}'}^{d+1}$ and let $\mathbf{p}^* \in A_{\mathbf{r}'}^{d+1}$ be the point such that $P' \subset K_{\mathbf{p}^*}$. We will also use the following notation: For each $\mathbf{p} \in A_{\mathbf{r}'}^{d+1}$, we will use $\mathbf{p}[m_1, m_2]$ to denote the point $\mathbf{q} \subset \mathbb{Z}^{d+1}$ such that $q_t = m_1$, $q_{d+1} = m_2$ and $q_i = p_i, \forall i : 1 \leq i \neq t \leq d$.

**Statement 1.** *If $p_t^* = 0$, then $p_{d+1}^* = 4b$ and furthermore, for each $\mathbf{p} \in P'$ such that Color$_{C'}$ [p] $\neq d+1$, Color$_C$ [$\psi(\mathbf{p}[p_t, 4b+1])$] = Color$_{C'}$ [p].*

*Proof.* Note first $p_{d+1}^* \neq 4b+1$, for otherwise, $K_{\mathbf{p}^*}$ does not contain color $d+1$. Second, if $p_{d+1}^* < 4b$, then each point in $\mathbf{q} \in K_{\mathbf{p}^*}$ is colored according one of the conditions in line 3, 8 or 10 of figure 4. Let $\mathbf{q}^* \in K_{\mathbf{p}^*}$ be the "red" point in $P'$. Then $\mathbf{q}^*$ must satisfy the condition in line 6 and hence there exists $l$ such that $q_l^* = r_l' - 1$. By our assumption, $p_t^* = 0$. Thus if $p_{d+1}^* < 4b$, then $l \notin \{t, d+1\}$, implying for each $\mathbf{q} \in K_{\mathbf{p}^*}$, $q_l > 0$ (as $q_l \geq q_l^* - 1 > 0$) and $\text{Color}_{C'}[\mathbf{q}] \neq l$. So $K_{\mathbf{p}^*}$ does not contain color $l$, contradicting with the assumption of the statement. Putting these two cases together, we have $p^*_{d+1} = 4b$.

We now prove the second part of the statement. If $p_{d+1} = 4b+1$, then we are done, because $\text{Color}_C[\psi(\mathbf{p})] = \text{Color}_{C'}[\mathbf{p}]$ according to line 1 of Figure 4. So, let us assume $p_{d+1} = 4b$. Because the statement assumes $\text{Color}_{C'}[\mathbf{p}] \neq d+1$, $\mathbf{p}$ satisfies the condition in line 3 and hence $\mathbf{p} \in \partial\left(A^{d+1}_{\mathbf{r}'}\right)$. By Property 1, we have $\text{Color}_{C'}[\mathbf{p}[p_t, 4b+1]] = \text{Color}_{C'}[\mathbf{p}]$, completing the proof of the statement. □

**Statement 2.** If $p^*_t = a+2$ or $a+3$, then $p^*_{d+1} = 0$, and in addition, for each $\mathbf{p} \in P'$ such that $\text{Color}_{C'}[\mathbf{p}] \neq d+1$, $\text{Color}_C[\psi(\mathbf{p}[p_t, 1])] = \text{Color}_{C'}[\mathbf{p}]$.

*Proof.* If $p^*_{d+1} > 0$, then $K_{\mathbf{p}^*}$ does not contain color $d+1$. So $p^*_{d+1} = 0$. In this case, $p_{d+1}$ must be 1, since $\text{Color}_{C'}[\mathbf{q}] = d+1$ for any $\mathbf{q} \in A^{d+1}_{\mathbf{r}'}$ with $q_{d+1} = 0$. Thus, $\text{Color}_C[\psi(\mathbf{p}[p_t, 1])] = \text{Color}_{C'}[\mathbf{p}[p_t, 1]] = \text{Color}_{C'}[\mathbf{p}]$. □

**Statement 3.** If $p^*_{d+1} = 4b$, then $0 \leq p^*_t \leq a+1$, and moreover, for each $\mathbf{p} \in P'$ such that $\text{Color}_{C'}[\mathbf{p}] \neq d+1$, $\text{Color}_C[\psi(\mathbf{p}[p_t, 4b+1])] = \text{Color}_{C'}[\mathbf{p}]$.

*Proof.* The first part of the statement is straightforward. Similar to the proof of **Statement 1**, we can prove the second part of the statement for the case that $0 \leq p_t \leq a+1$. When $p_t = a+2$, we have $\psi(\mathbf{p}) = \psi(\mathbf{p}[p_t, 4b+1])$. Thus, $\text{Color}_C[\psi(\mathbf{p}[p_t, 4b+1])] = \text{Color}_C[\psi(\mathbf{p})] = \text{Color}_{C'}[\mathbf{p}]$. □

We can similarly prove the following statements.

**Statement 4.** If $p^*_{d+1} = 4i+1$ or $4i+2$ for some $0 \leq i \leq b-1$, then $p^*_t = 1$ and furthermore, for each $\mathbf{p} \in P'$ such that $\text{Color}_{C'}[\mathbf{p}] \neq d+1$, $\text{Color}_C[\psi(\mathbf{p}[2, p_{d+1}])] = \text{Color}_{C'}[\mathbf{p}]$.

**Statement 5.** If $p^*_{d+1} = 4i$ for some $1 \leq i \leq b-1$, then $1 \leq p^*_t \leq a+1$, and in addition, for each $\mathbf{p} \in P'$ such that $\text{Color}_{C'}[\mathbf{p}] \neq d+1$, if $2 \leq p_t \leq a+1$, then $\text{Color}_C[\psi(\mathbf{p}[p_t, 4i+1])] = \text{Color}_{C'}[\mathbf{p}]$, and if $p_t = 1$, then $\text{Color}_C[\psi(\mathbf{p}[2, 4i+1])] = \text{Color}_{C'}[\mathbf{p}]$.

**Statement 6.** If $p^*_{d+1} = 4i-1$ for some $1 \leq i \leq b$, then $1 \leq p^*_t \leq a+1$, and moreover, for each $\mathbf{p} \in P'$ such that $\text{Color}_{C'}[\mathbf{p}] \neq d+1$, if $2 \leq p_t \leq a+1$, then $\text{Color}_C[\psi(\mathbf{p}[p_t, 4i-1])] = \text{Color}_{C'}[\mathbf{p}]$, and if $p_t = 1$, then $\text{Color}_C[\psi(\mathbf{p}[2, 4i-1])] = \text{Color}_{C'}[\mathbf{p}]$.

**Statement 7.** If $p^*_{d+1} = 0$, then $1 \leq p^*_t \leq a+3$, and in addition, for each $\mathbf{p} \in P'$ such that $\text{Color}_{C'}[\mathbf{p}] \neq d+1$, if $2 \leq p^*_t \leq a+1$, then $\text{Color}_C[\psi(\mathbf{p}[p_t, 1])] = \text{Color}_{C'}[\mathbf{p}]$, and if $p^*_t = 1$, then $\text{Color}_C[\psi(\mathbf{p}[2, 1])] = \text{Color}_{C'}[\mathbf{p}]$.

In addition,

**Statement 8.** $p^*_{d+1} \neq 4b+1$.

*Proof.* This statement is true if $p^*_{d+1} = 4b+1$ then $K_{\mathbf{p}^*}$ does not contain color $d+1$. □

Now suppose that $P'$ is a panchromatic simplex of $T'$. Let $\mathbf{p}^*$ be the point such that $P' \subset K_{\mathbf{p}^*}$. Then $P'$ and $\mathbf{p}^*$ must satisfy the conditions of one of the the statements. By that statement, we can transform every point $\mathbf{p} \in P'$, except for the one that has color $d+1$, back to a point $\mathbf{q}$ in $A_{\mathbf{r}}^d$ to obtain a set $P$ from $P'$. This transformation maintains the coloring. As $P$ consists of the $d+1$ points and is accommodated, it is a panchromatic simplex of $C$. Thus, with all the statements above, we specify an efficient algorithm to compute a panchromatic simplex $P$ of $T$ given a panchromatic simplex $P'$ of $T'$. □

## 3.2 PPAD-Completeness of Problem BROUWER$^f$

We are now ready to prove the main result of this section.

**Theorem 1 (High Dimensional Brouwer's Fixed Points).** *For any well-behaved function $f$, search problem* BROUWER$^f$ *is* **PPAD**-*complete.*

*Proof.* We reduce BROUWER$^{f_2}$ to BROUWER$^f$ in order to prove the latter is **PPAD**-complete. Recall, $f_2(n) = \lfloor n/2 \rfloor$. Suppose $(C, 0^{2n})$ is an input instance of BROUWER$^{f_2}$. Let

$$l = f(11n) \geq 3, \quad m' = \left\lceil \frac{n}{l-2} \right\rceil \quad \text{and} \quad m = \left\lceil \frac{11n}{l} \right\rceil.$$

We iteratively construct a sequence of coloring triples $\mathcal{T} = \{\, T^0, T^1, \ldots T^{w-1}, T^w \,\}$ for some $w = O(m)$, starting with $T^0 = (C, 2, (2^n, 2^n))$ and ending with $T^w = (C^w, m, \mathbf{r}^w)$ where $\mathbf{r}^w \in \mathbb{Z}^m$ and $\mathbf{r}_i^w = 2^l$, $\forall i : 1 \leq i \leq m$. At the $t^{th}$, we apply either $\mathbf{L}^1, \mathbf{L}^2$ or $\mathbf{L}^3$ with properly chosen parameters to build $T^{t+1}$ from $T^t$.

Below we give the details of our construction. In the first step, we call $\mathbf{L}^1\left(T^0, 1, 2^{m'(l-2)}\right)$ to get $T^1 = \left(C^1, 2, \left(2^{m'(l-2)}, 2^n\right)\right)$. This step is possible because $m'(l-2) \geq n$. We then invoke the procedure in Figure 5. In each for-loop, the first component of $\mathbf{r}$ decreases by a factor of $2^{l-2}$, while the dimension of the space increases by 1. After running the for-loop $(m'-5)$ times, we obtain a coloring triple $T^{3m'-14} = \left(C^{3m'-14}, d^{3m'-14}, \mathbf{r}^{3m'-14}\right)$ that satisfies[6]

$$d^{3m'-14} = m' - 3, \quad r_1^{3m'-14} = 2^{5(l-2)}, \quad r_2^{3m'-14} = 2^n \text{ and } r_i^{3m'-14} = 2^l, \ \forall\, i : 3 \leq i \leq m' - 3.$$

Next, we call the procedure given in Figure 6. Note that the while-loop must terminate in at most 8 iterations because we start with $r_1^{3m'-14} = 2^{5(l-2)}$. The procedure returns a coloring triple $T^{w'} = \left(C^{w'}, d^{w'}, \mathbf{r}^{w'}\right)$ that satisfies

$$w' \leq 3m' + 11, \quad d^{w'} \leq m' + 5, \quad r_1^{w'} = 2^l, \quad r_2^{w'} = 2^n \text{ and } r_i^{w'} = 2^l, \ \forall\, i : 3 \leq i \leq d^{w'}.$$

Then we repeat the whole process above on the second coordinate and obtain a coloring triple $T^{w''} = \left(C^{w''}, d^{w''}, \mathbf{r}^{w''}\right)$ that satisfies

$$w'' \leq 6m' + 21, \quad d^{w''} \leq 2m' + 8 \text{ and } r_i^{w''} = 2^l, \forall\, i : 1 \leq i \leq d^{w''}.$$

---

[6]**Remark: the superscript of $C$, $d$, $r_i$, denotes the index of the iterative step, it is not an exponent!.**

**The Construction of $T^{3m'-14}$ from $T^1$**

1: **for any** $t$ from 0 to $m'-6$ **do**
2:     It can be proved inductively that $T^{3t+1} = (C^{3t+1}, d^{3t+1}, \mathbf{r}^{3t+1})$ satisfies
$$d^{3t+1} = t+2,\ r_1^{3t+1} = 2^{(m'-t)(l-2)},\ r_2^{3t+1} = 2^n \text{ and } r_i^{3t+1} = 2^l \text{ for any } 3 \leq i \leq t+2$$
3:     let $u = (2^{(m'-t-1)(l-2)} - 5)(2^{l-1} - 1) + 5$
4:     [ $u \geq r_1^{3t+1} = 2^{(m'-t)(l-2)}$ under the assumption that $t \leq m' - 6$ and $l \geq 3$ ]
5:     $T^{3t+2} = \mathbf{L}^1(T^{3t+1}, 1, u)$
6:     $T^{3t+3} = \mathbf{L}^3(T^{3t+2}, 1, 2^{(m'-t-1)(l-2)}, 2^{l-2} - 1)$
7:     $T^{3t+4} = \mathbf{L}^1(T^{3t+3}, t+3, 2^l)$

Figure 5: The Construction of $T^{3m'-14}$ from $T^1$

**The Construction of $T^{w'}$ from $T^{3m'-14}$**

1: let $t = 0$
2: **while** $T^{3(m'+t)-14} = (C^{3(m'+t)-14}, m'+t-3, \mathbf{r}^{3(m'+t)-14})$ satisfies $r_1^{3(m'+t)-14} > 2^l$ **do**
3:     let $k = \lceil (r_1^{3(m'+t)-14} - 5)/(2^{l-1} - 1) \rceil + 5$
4:     $T^{3(m'+t)-13} = \mathbf{L}^1(T^{3(m'+t)-14}, 1, (k-5)(2^{l-1}-1)+5)$
5:     $T^{3(m'+t)-12} = \mathbf{L}^3(T^{3(m'+t)-13}, 1, k, 2^{l-2} - 1)$
6:     $T^{3(m'+t)-11} = \mathbf{L}^1(T^{3(m'+t)-12}, m'+t-2, 2^l)$, set $t = t+1$
8: let $w' = 3(m'+t)-13$ and $T^{w'} = \mathbf{L}^1(T^{3(m'+t)-14}, 1, 2^l)$

Figure 6: The Construction of $T^{w'}$ from $T^{3m'-14}$

The way we define $m$ and $m'$ guarantees

$$d^{w''} \leq 2m' + 8 \leq 2\left(\frac{n}{l-2} + 1\right) + 8 \leq 2\left(\frac{n}{l/3}\right) + 10 = \frac{6n}{l} + 10 \leq \frac{11n}{l} \leq m.$$

Finally, by applying $\mathbf{L}^2$ for $m - d^{w''}$ times with parameter $u = 2^l$, we obtain $T^w = (C^w, m, \mathbf{r}^w)$ with $r_i^w = 2^l, \forall 1 \leq i \leq m$. It follows from our construction, $w = O(m)$.

To see why the sequence $\mathcal{T}$ gives a reduction from BROUWER$^{f_2}$ to BROUWER$^f$, let $T^i = (C^i, d^i, \mathbf{r}^i)$ (again the superscript of $C$, $d$, $r_i$, denotes the index of the iteration). As sequence $\{\text{Size}[\mathbf{r}^i]\}_{0 \leq i \leq w}$ is nondecreasing and $w = O(m) = O(n)$, by Property **A** of Lemma 2, 3 and 4, there exists a polynomial $g(n)$ such that

$$\text{Size}[C^w] = \text{Size}[C] + O(g(n)).$$

By these Properties **A** again, we can construct the whole sequence $\mathcal{T}$ and in particular, $T^w = (C^w, m, \mathbf{r}^2)$, in time polynomial in $\text{Size}[C]$.

$(C^w, 0^{11n})$ is an input instance of BROUWER$^f$. Given a panchromatic simplex $P$ of $(C^w, 0^{11n})$, using the algorithm in Property **B** of Lemma 2, 3 and 4, we can compute a sequence of panchromatic simplex $P^w = P, P^{w-1}..., P^0$ iteratively in polynomial time, where $P^t$ is a panchromatic simplex of $T^t$ and is computed from the panchromatic simplex $P^{t+1}$ of $T^{t+1}$. In the end, we obtain $P^0$, which is a panchromatic set of $(C, 0^{2n})$. □

## 4 Hardness of Approximating Nash Equilibria

In this section, we show it is unlikely that the bimatrix game has a fully polynomial-time approximation scheme. More precisely, we reduce BROUWER$^{f_1}$, for $f_1(n) = 3$, to the computation of a $1/n^{\Theta(1)}$-approximate Nash equilibrium of a bimatrix game $(\mathbf{A}, \mathbf{B})$ where $\mathbf{A}, \mathbf{B} \in \mathbb{R}_{[0,1]}^{n \times n}$. Thus, the latter problem is also **PPAD**-complete.

### 4.1 Basic Notations and the Main Results

We start with some notations that are useful for our reduction and analysis.

#### 4.1.1 $\epsilon$-Well-Supported Nash Equilibria

A key concept in the reduction of Daskalakis, Goldberg and Papadimitriou [7] and Chen and Deng [6] is an alternative notion of approximate Nash equilibria. To distinguish it from the more commonly used notion of approximate Nash equilibria, we will refer to this alternative approximation as an $\epsilon$-well-supported Nash equilibrium.

Let $\mathcal{G} = (\mathbf{A}, \mathbf{B})$ be a bimatrix game where $\mathbf{A}$ and $\mathbf{B}$ are two $n \times n$ matrices. We use $\mathbf{a}_i$ to denote the $i^{th}$ row of $\mathbf{A}$ and $\mathbf{b}_i$ to denote the $i^{th}$ column of $\mathbf{B}$. In a profile of mixed strategies $(\mathbf{x}, \mathbf{y})$, the payoff of the first player if he/she chooses the $i^{th}$ strategy is $\langle \mathbf{a}_i | \mathbf{y} \rangle$, and the payoff of the second player if he/she chooses the $i^{th}$ strategy is $\langle \mathbf{b}_i | \mathbf{x} \rangle$.

**Definition 7 ($\epsilon$-well-supported Nash Equilibria).** *An $\epsilon$-well-supported Nash equilibrium of a bimatrix game $(\mathbf{A}, \mathbf{B})$ is a profile of mixed strategies $(\mathbf{x}^*, \mathbf{y}^*)$, such that for all $1 \leq i, j \leq n$,*

$$\langle \mathbf{b}_i | \mathbf{x}^* \rangle > \langle \mathbf{b}_j | \mathbf{x}^* \rangle + \epsilon \Rightarrow y_j^* = 0 \quad and \quad \langle \mathbf{a}_i | \mathbf{y}^* \rangle > \langle \mathbf{a}_j | \mathbf{y}^* \rangle + \epsilon \Rightarrow x_j^* = 0.$$

Recall that the commonly used notion of approximation is defined as:

**Definition 8 ($\epsilon$-approximate Nash equilibria).** *An $\epsilon$-approximate Nash equilibrium of game $(\mathbf{A}, \mathbf{B})$ is a profile of mixed strategies $(\mathbf{x}^*, \mathbf{y}^*)$, such that for all probability vectors $\mathbf{x}, \mathbf{y} \in \mathbb{P}^n$,*

$$(\mathbf{x}^*)^T \mathbf{A} \mathbf{y}^* \geq \mathbf{x}^T \mathbf{A} \mathbf{y}^* - \epsilon \quad and \quad (\mathbf{x}^*)^T \mathbf{B} \mathbf{y}^* \geq (\mathbf{x}^*)^T \mathbf{B} \mathbf{y} - \epsilon.$$

We now show that these two notions are polynomially related. This polynomial relation allows us to prove the **PPAD** result with a pair-wise approximation condition. Thus, we can locally argue certain properties of the bimatrix game that we build from the fixed-point problem.

**Lemma 5 (Polynomially equivalence of the two notions of approximate Nash Equilibria).**

1. *From any $\epsilon^2/(8n)$-approximate Nash equilibrium $(\mathbf{u}, \mathbf{v})$ of game $(\mathbf{A}, \mathbf{B})$, we can compute in polynomial time an $\epsilon$-well-supported Nash equilibrium $(\mathbf{x}, \mathbf{y})$ of $(\mathbf{A}, \mathbf{B})$.*

2. *For any $0 \leq \epsilon \leq 1$ and for any bimatrix game $(\mathbf{A}, \mathbf{B})$ where $\mathbf{A}, \mathbf{B} \in \mathbb{R}^{n \times n}_{[0:1]}$, if $(\mathbf{x}, \mathbf{y})$ is an $\epsilon$-well-supported Nash equilibrium of $(\mathbf{A}, \mathbf{B})$, then $(\mathbf{x}, \mathbf{y})$ is also an $\epsilon$-approximate Nash equilibrium of $(\mathbf{A}, \mathbf{B})$.*

*Proof.* By the definition of approximate Nash equilibria, we have

$$\forall\, \mathbf{u}' \in \mathbb{P}^n, \qquad (\mathbf{u}')^T \mathbf{A} \mathbf{v} \leq \mathbf{u}^T \mathbf{A} \mathbf{v} + \epsilon^2/(8n),$$
$$\forall\, \mathbf{v}' \in \mathbb{P}^n, \qquad \mathbf{u}^T \mathbf{B} \mathbf{v}' \leq \mathbf{u}^T \mathbf{B} \mathbf{v} + \epsilon^2/(8n).$$

Consider some $j$ with some $i$ such that $\langle \mathbf{a}_i | \mathbf{v} \rangle \geq \langle \mathbf{a}_j | \mathbf{v} \rangle + \epsilon/2$, where $\mathbf{a}_i$ is the $i^{th}$ row of matrix $\mathbf{A}$. By changing $u_j$ to 0 and $u_i$ to $u_i + u_j$ we can increase the first-player's profit by $u_j(\epsilon/2)$, implying $u_j < \epsilon/(4n)$. Similarly, all such $j$ have $v_j < \epsilon/(4n)$.

We now set all these $u_j$ and $v_j$ to 0 and uniformly increase the probability of other strategies to obtain a new pair of mixed strategies, $(\mathbf{x}, \mathbf{y})$.

Note for all $i$, $|\langle \mathbf{a}_i | \mathbf{x} \rangle - \langle \mathbf{a}_i | \mathbf{u} \rangle| \leq \epsilon/4$, because we assume the absolute value of each entry in $\mathbf{a}_i$ is less then 1. Thus, the relative change between $\langle \mathbf{a}_i | \mathbf{x} \rangle$ and $\langle \mathbf{a}_j | \mathbf{x} \rangle$ is no more than $\epsilon/2$. Thus, any $j$ that is beaten some $i$ by a gap of $\epsilon$ is set to zero in $(\mathbf{x}, \mathbf{y})$.

Part 2 follows directly from the definitions. $\square$

### 4.1.2 Problem BROUWER

To simplify the presentation of our reduction, we introduce a problem called BROUWER that is equivalent to BROUWER$^{f_1}$ for $f_1(n) = 3$.

For any $n \in \mathbb{Z}^+$, let $B^n = \mathbb{Z}^n_{[0,7]}$.

**Definition 9 (BROUWER).** *The input instance of BROUWER is a pair $(C, 0^n)$ where $C$ is a valid Brouwer-mapping circuit with parameters $n$ and $\mathbf{r}$ where $r_i = 8$ for any $i : 1 \leq i \leq n$. $C$ specifies a coloring assignment $\text{Color}_C : B^n \to \{1, 2, \ldots n+1\}$ in the same way as in definition 5. The only difference is that we use color $d+1$ to encode the special color "red".*

*Recall that a set $P \subset B^n$ is panchromatic set of a panchromatic simplex if it is accommodated, has all $(n+1)$ colors, and $|P| = n+1$.*

*The output of the this search problem is a panchromatic simplex $P$.*

Clearly, BROUWER is equivalent to BROUWER$^{f_1}$, and thus is **PPAD**-complete.

### 4.1.3 The main result

In this section, we consider a bimatrix game $(\mathbf{A}, \mathbf{B})$ where $\mathbf{A}, \mathbf{B} \in \mathbb{R}^n_{[0,1]}$. We call such a game $(\mathbf{A}, \mathbf{B})$ a *positively normalized* bimatrix game. Although we model the entries of the bimatrix games as real numbers, our results extend to the case when all entries are between 0 and 1, and are given in binary representations.

Our main result of this section can be stated as:

**Theorem 2 (Main).** *The problem of computing a $1/n^6$-well-supported Nash equilibrium of a positively normalized $n \times n$ bimatrix game is* **PPAD**-*complete*.

Together with Lemma 5, we can prove the following theorem which implies that the second conjecture of the Introduction is not likely to be true, unless **PPAD** $\subset$ **FP**.

**Theorem 3 (Unlikely Fully Polynomial-Time Approximation).** *The problem of computing a $1/n^{\Theta(1)}$-approximate Nash equilibrium of a positively normalized $n \times n$ bimatrix game is* **PPAD**-*complete*.

## 4.2 Outline of the Reduction

Our reduction is built upon the earlier work of Daskalakis, Goldberg and Papadimitriou [7] and Chen and Deng [6]. In particular, we extend the construction of Chen and Deng [6] to Brouwer's fixed-point problem in high dimensions. The consideration of high dimensional discrete fixed point problems is crucial to our improvement of the approximation ratio from $2^{-\Theta(n)}$ to $1/n^{\Theta(1)}$. Naturally, the reduction from high dimensional problems introduces new technical challenges, and we cannot just naively extend the earlier construction. For example, we develop a new averaging-maneuvering scheme to overcome the curse of high dimensionality.

We would like to point our that we are not optimizing the size of the bimatrix game in our reduction at the expense of the simplicity of the presentation. The only objective here is to construct a bimatrix game whose size is polynomial in Size $[C]$. Recall Size $[C]$ is the number of gates plus the number of input and output variables in $C$. Note that $n <$ Size $[C]$, as $C$ has $2n$ output bits.

**The Bimatrix Games**

Let $(C, 0^n)$ be an input instance of problem BROUWER. We first construct a bimatrix game $\mathcal{G} = (\mathbf{A}, \mathbf{B})$ in polynomial time. Both players in the game have $N = 2^{6m+1} = 2K$ strategies where $m$ is the smallest integer such that $2^m \geq$ Size $[C]$. Our construction will ensure that:

- **Property $\mathbf{A}_1$**: $|a_{i,j}|, |b_{i,j}| \leq N^3$ for all $i, j$: $1 \leq i, j \leq N$;

- **Property $\mathbf{A}_2$**: From each $\epsilon$-well-supported Nash equilibrium of $\mathcal{G}$, where $\epsilon = 2^{-18m} = 1/K^3$, we can compute a panchromatic simplex $P$ of circuit $C$ in polynomial time.

We then transform $\mathcal{G}$ into a positively normalized bimatrix game $\bar{\mathcal{G}} = (\bar{\mathbf{A}}, \bar{\mathbf{B}})$ by setting

$$\bar{a}_{i,j} = \frac{a_{i,j} + N^3}{2N^3} \quad \text{and} \quad \bar{b}_{i,j} = \frac{b_{i,j} + N^3}{2N^3}$$

for all $i, j : 1 \leq i, j \leq N$. Using property $\mathbf{A}_2$, we can compute a panchromatic simplex $P$ of $C$ in polynomial time from any $N^{-6}$-well-supported Nash equilibrium[7] of game $\bar{\mathcal{G}}$.

In the remainder of this section, we set $\epsilon = 2^{-18m} = 1/K^3$ and recall $K = 2^{6m}$ and $N = 2K$.

---
[7]Actually, a $4N^{-6}$-well-supported Nash equilibrium would be sufficient.

### The Strategies and Arithmetic Networks

Let us denote the two players by $P_1$ and $P_2$. In the reduction, we will build an arithmetic network to model the circuit $C$ and the condition of a Brouwer's fixed point. In this network, there are two sets of nodes:

$V_A$: the set of $K = 2^{6m}$ *arithmetic nodes*, and $V_I$: the set of $K = 2^{6m}$ *internal nodes*.

We always use $v$ to refer to a node in $V_A$ and $w$ to refer to a node in $V_I$. Each arithmetic node or internal node contributes two pure strategies, a boolean 0-strategy and a boolean 1-strategy, to the bimatrix game $\mathcal{G}$. Arbitrarily, we pick a one-to-one correspondence $\mathcal{C}_A$ from $V_A$ to $\{1, 2, ..., n\}$, and a one-to-one correspondence $\mathcal{C}_I$ from $V_I$ to $\{1, 2, ..., n\}$. Every $v \in V_A$ contributes to the $2\mathcal{C}_A(v) - 1^{st}$ and $2\mathcal{C}_A(v)^{th}$ row strategies of $\mathcal{G}$. Similarly, $w \in N_I$ contributes to the $2\mathcal{C}_I(w) - 1^{st}$ and $2\mathcal{C}_I(w)^{th}$ column strategies of $\mathcal{G}$. Recall that each player has $N = 2K$ strategies.

Suppose we have somehow assembled game $\mathcal{G} = (\mathbf{A}, \mathbf{B})$. Let $(\mathbf{x}, \mathbf{y})$ be a mixed-strategy profile of $\mathcal{G}$. We will let $\mathbf{x}[v] = x_{2\mathcal{C}_A(v)-1}$ to denote the probability that the first player $P_1$ chooses row $2\mathcal{C}_A(v) - 1$ and let $\mathbf{x}_C[v] = x_{2\mathcal{C}_A(v)-1} + x_{2\mathcal{C}_A(v)}$ to denote the probability that $P_1$ chooses either row $2\mathcal{C}_A(v) - 1$ or row $2\mathcal{C}_A(v)$. Similarly, we let $\mathbf{y}[w]$ to denote the probability that the second player $P_2$ chooses column $2\mathcal{C}_I(w) - 1$ and $\mathbf{y}_C[w]$ to denote the probability $P_2$ chooses either column $2\mathcal{C}_I(w) - 1$ or column $2\mathcal{C}_I(w)$. We refer to $\mathbf{x}[v]$ and $\mathbf{y}[w]$ as the *values* of these nodes, and $\mathbf{x}_C[v]$ and $\mathbf{y}_C[w]$ as the *capacities* of these nodes.

Following [7, 6], given an $\epsilon$-well-supported Nash equilibrium $(\mathbf{x}, \mathbf{y})$ of $\mathcal{G}$, we view $\mathbf{x}[v]$ as a meaningful real number. We use the same set of nine arithmetic and logic gadgets designed in [6] to build the game $\mathcal{G}$ that models BROUWER with instance $(C, 0^n)$. Every gadget contains exactly one interior node in $V_I$ to mediate between arithmetic nodes in the gadget in order to ensure that the values of the latter ones obey the intended arithmetic or logic relationship.

### Building from the Zero-Sum Penny Matching Game

To construct $\mathcal{G}$ we start with a bimatrix game $\mathcal{G}^* = (\mathbf{A}^*, \mathbf{B}^*)$ called *Matching Pennies* with payoff parameter $M = 2^{18m+1} = 2K^3$. Each player in $\mathcal{G}^*$ has same intended number of strategies as those in $\mathcal{G}$, thus, $\mathbf{A}^*$ and $\mathbf{B}^*$ are $N \times N$ matrices, where recall $N = 2^{6m+1}$. Their entries are chosen from $\{0, M, -M\}$ and are specified according to Figure 7.

To construct $\mathcal{G}$, we add a polynomial number of gadgets into the prototype game $\mathcal{G}^*$ to form a network over arithmetic nodes $V_A$ that models problem BROUWER with instance $(C, 0^n)$ in order to guarantee Property $\mathbf{A}_2$. Each gadget contains exactly one interior node in $V_I$ and no more than three arithmetic nodes in $V_A$. One of the arithmetic nodes is refer to as the *output node* of the gadget and others are the *input nodes*.

Suppose $w$ is the interior node and $v$ is the output arithmetic node of a gadget $G$. By saying inserting $G$ into the network and hence the game $\mathcal{G}^*$, we modify the following payoff entries of $\mathcal{G}^*$ related to nodes $v$ and $w$:

$$\text{the } 2\mathcal{C}_A(v) - 1^{st} \text{ and } 2\mathcal{C}_A(v)^{th} \text{ rows of } \mathbf{A}^*;$$
$$\text{the } 2\mathcal{C}_I(w) - 1^{st} \text{ and } 2\mathcal{C}_I(w)^{th} \text{ columns of } \mathbf{B}^*.$$

**Matching Pennies with Payoff M**

1: **for** any $i, j : 1 \leq i, j \leq K = 2^{6m}$ **do**
2:     **if** $i = j$ **then**
3:         set $a^*_{2i-1,2j-1} = a^*_{2i-1,2j} = a^*_{2i,2j-1} = a^*_{2i,2j} = M$
4:         set $b^*_{2i-1,2j-1} = b^*_{2i-1,2j} = b^*_{2i,2j-1} = b^*_{2i,2j} = -M$
5:     **else**
6:         set $a^*_{2i-1,2j-1} = a^*_{2i-1,2j} = a^*_{2i,2j-1} = a^*_{2i,2j} = 0$
7:         set $b^*_{2i-1,2j-1} = b^*_{2i-1,2j} = b^*_{2i,2j-1} = b^*_{2i,2j} = 0$

Figure 7: Matrices $\mathbf{A}^*$ and $\mathbf{B}^*$ of Game $\mathcal{G}^*$

To insert $G$, we add constants in $[0 : 1]$ to these two rows of $\mathbf{A}^*$ and these two columns of $\mathbf{B}^*$. Each type of gadgets has its own constants. During the construction of the arithmetic network and the game $\mathcal{G}$, as each arithmetic node in $V_A$ is used as an output node in at most one gadget (although it is allowed to be used as an input node in arbitrarily many gadgets) and each interior node in $V_I$ is used in at most one gadget, every entry in $\mathbf{A}^*$ and $\mathbf{B}^*$ is modified at most once.

In the remainder of this section, we give the details of $\mathcal{G}$ and prove the correctness of our reduction.

### 4.3 Properties of the Prototype Game $\mathcal{G}^*$

We start with an important property of the prototype game $\mathcal{G}^*$. As we have discussed in the previous subsection, in our reduction, we modify each entry of $\mathcal{G}^*$ at most once with a constant between 0 and 1. Let us define a class $\mathcal{L}$ containing all these intermediate games.

**Definition 10 (Class $\mathcal{L}$).** *A bimatrix game $\mathcal{G}' = (\mathbf{A}', \mathbf{B}')$ belongs to class $\mathcal{L}$ if $\mathbf{A}'$ and $\mathbf{B}'$ are two $N \times N$ matrices satisfying*

$$a^*_{i,j} \leq a'_{i,j} \leq a^*_{i,j} + 1 \quad \text{and} \quad b^*_{i,j} \leq b'_{i,j} \leq b^*_{i,j} + 1,$$

*for all $i, j : 1 \leq i, j \leq N$.*

A well-known fact about the game $\mathcal{G}^*$ is that, letting $(\mathbf{x}^*, \mathbf{y}^*)$ be a Nash equilibrium of $\mathcal{G}^*$, $x^*_C[v] = y^*_C[w] = 1/K$, for any $v \in V_A$ and $w \in V_I$. We now prove that, for any bimatrix game $\mathcal{G}' \in \mathcal{L}$, both players choose nodes (nearly) uniformly, in every $t$-well-supported Nash equilibrium for $t \leq 1$.

**Lemma 6 (Nearly Uniform Capacities).** *For any $t \leq 1$, let $(\mathbf{x}, \mathbf{y})$ be a $t$-well-supported Nash equilibrium of game $(\mathbf{A}, \mathbf{B}) \in \mathcal{L}$. Then for all $v \in V_A$ and $w \in V_I$, their capacities satisfy*

$$\left| \mathbf{x}_C[v] - 1/K \right| < \epsilon = 2^{-18m} \quad \text{and} \quad \left| \mathbf{y}_C[w] - 1/K \right| < \epsilon = 2^{-18m}.$$

*Proof.* By the definition of Class $\mathcal{L}$, for each $k$, the $2k-1^{st}$ and $2k^{th}$ entries of rows $\mathbf{a}_{2k-1}$ and $\mathbf{a}_{2k}$ in $\mathbf{A}$ are within $[M : M+1]$ and all other entries in them are within $[0 : 1]$. Thus, for any probability vector $\mathbf{y} \in \mathbb{P}^n$, if the node $w \in V_I$ has $\mathcal{C}_I(w) = k$, then

$$M\mathbf{y}_C[w] \leq \langle \mathbf{a}_{2k-1} | \mathbf{y} \rangle \leq M\mathbf{y}_C[w] + 1 \quad \text{and} \quad M\mathbf{y}_C[w] \leq \langle \mathbf{a}_{2k} | \mathbf{y} \rangle \leq M\mathbf{y}_C[w] + 1 \qquad (1)$$

Similarly, for each $l$, the $2l - 1^{st}$ and $2l^{th}$ entries of columns $\mathbf{b}_{2k-1}$ and $\mathbf{b}_{2k}$ in $\mathbf{B}$ are within $[-M : -M+1]$ and all other entries in them are within $[0 : 1]$. Thus, for any probability vector $\mathbf{x} \in \mathbb{P}$, if node $v \in V_A$ has $\mathcal{C}_A(v) = l$, then

$$-M\mathbf{x}_C[v] \leq \langle \mathbf{b}_{2l-1} | \mathbf{x} \rangle \leq -M\mathbf{x}_C[v] + 1 \quad \text{and} \quad -M\mathbf{x}_C[v] \leq \langle \mathbf{b}_{2l} | \mathbf{x} \rangle \leq -M\mathbf{x}_C[v] + 1. \qquad (2)$$

Now suppose $(\mathbf{x}, \mathbf{y})$ be a $t$-well-supported Nash equilibrium of $(\mathbf{A}, \mathbf{B})$ for $t \leq 1$. To warm up, we first prove that for each pair of $v$ and $w$ with $\mathcal{C}_A(v) = \mathcal{C}_I(w)$, say they are equal to $l$, if $\mathbf{y}_C[w] = 0$ then $\mathbf{x}_C[v] = 0$. First not that $\mathbf{y}_C[w] = 0$ implies that there exists a node $w' \in V_I$ with capacity $\mathbf{y}_C[w'] \geq 1/K$. Suppose $\mathcal{C}_I(w) = k$. By Inequality (1),

$$\langle \mathbf{a}_{2k} | \mathbf{y} \rangle - \max\left( \langle \mathbf{a}_{2l} | \mathbf{y} \rangle, \langle \mathbf{a}_{2l-1} | \mathbf{y} \rangle \right) \geq M\mathbf{y}_C[w'] - (M\mathbf{y}_C[w] + 1) \geq M/K - 1 > 1$$

In other words, the payoff of $P_1$ with the choice of the $2k^{th}$ strategy is more than 1 plus the the payoff if $P_1$ chooses the $2l^{th}$ or the $2l - 1^{st}$ strategy. Because $(\mathbf{x}, \mathbf{y})$ is a $t$-well-supported Nash equilibrium with $t \leq 1$, we have $\mathbf{x}_C[v] = 0$.

Next, we prove $\left| \mathbf{x}_C[v] - 1/K \right| < \epsilon$ for all $v \in V_A$. To derive a contradiction, we assume this statement is not true. Then there exist $v, v' \in N_1$ such that $\mathbf{x}_C[v] - \mathbf{x}_C[v'] > \epsilon$. Let $l = \mathcal{C}_A(v)$ and $k = \mathcal{C}_A(v')$. By Inequality (2),

$$\langle \mathbf{b}_{2k} | \mathbf{x} \rangle - \max\left( \langle \mathbf{b}_{2l} | \mathbf{x} \rangle, \langle \mathbf{b}_{2l-1} | \mathbf{x} \rangle \right) \geq -M\mathbf{x}_C[v'] - (-M\mathbf{x}_C[v] + 1) > 1,$$

as $M = 2^{18m+1} = 2/\epsilon$. Thus this assumption would imply $\mathbf{y}_C[w] = 0$ for the node $w \in V_I$ with $\mathcal{C}_I(w) = l$, and in turn imply $\mathbf{x}_C[v] = 0$, which contradicts with our assumption that $\mathbf{x}_C[v] > \mathbf{x}_C[v'] + \epsilon > 0$.

We can similarly show $\left| \mathbf{y}_C[w] - 1/K \right| < \epsilon$ for all $w \in V_I$. □

## 4.4 Arithmetic and Logic Gadgets

We will use the nine $G_\zeta$, $G_{\times \zeta}$, $G_=$, $G_+$, $G_-$, $G_<$, $G_\wedge$, $G_\vee$ and $G_\neg$ designed by Chen and Deng [6] in our reduction. To be self-contained, we restate the definitions and the properties of these gadgets.

Among these nine gadgets, $G_\wedge$, $G_\vee$ and $G_\neg$ the logic gadgets. They will be used to simulate the gates in Boolean circuit $C$. As mentioned in [6], these gadgets perform properly only when the values of their input nodes are representations of binary bits.

Formally, associated with a mixed strategy profile $(\mathbf{x}, \mathbf{y})$, the value of node $v \in N_1$ represents boolean 1 if $\mathbf{x}[v] = \mathbf{x}_C[v]$; it encodes boolean 0 if $\mathbf{x}_C[v] = 0$.

Other gadgets will be used to perform necessary arithmetic and comparison operations. Below, we include the proof of the property of gadget $G_+$ to illustrate how such properties are established. Properties of others gadget can be established similarly.

For convenience, we abuse $v$ and $w$ to also denote integers $\mathcal{C}_A(v)$ and $\mathcal{C}_I(w)$. For example, by saying $2v$, we actually mean $2\mathcal{C}_A(v)$. This should be clear from the context. For any pure strategy profile $s = (i,j) \in \{1, 2, ..., N\}^2$, we let $a_s = a_{i,j}$ and $b_s = b_{i,j}$. Moreover, by $x = y \pm \epsilon$, we mean $y - \epsilon \leq x \leq y + \epsilon$.

**Proposition 1 (Gadget $G_+$).** *Let $\mathcal{G}' = (\mathbf{A}, \mathbf{B})$ be a bimatrix game in $\mathcal{L}$ and nodes $v_1, v_2, v_3 \in V_A$, $w \in V_I$. Let pure strategy profiles $s_1 = (2v_1 - 1, 2w - 1)$, $s_2 = (2v_2 - 1, 2w - 1)$, $s_3 = (2v_3 - 1, 2w)$, $s_4 = (2v_3 - 1, 2w - 1)$ and $s_5 = (2v_3, 2w)$. If $\mathcal{G}'$ satisfies*

1). $b_{s_1} = b^*_{s_1} + 1$, $b_{s_2} = b^*_{s_2} + 1$ and $b_{i,2w-1} = b^*_{i,2w-1}$, for any other $i : 1 \leq i \leq N$;

2). $b_{s_3} = b^*_{s_3} + 1$ and $b_{i,2w} = b^*_{i,2w}$, for any other $i : 1 \leq i \leq N$;

3). $a_{s_4} = a^*_{s_4} + 1$ and $a_{2v_3-1,i} = a^*_{2v_3-1,i}$, for any other $i : 1 \leq i \leq N$;

4). $a_{s_5} = a^*_{s_5} + 1$ and $a_{2v_3,i} = a^*_{2v_3,i}$, for any other $i : 1 \leq i \leq N$,

*then in any $\epsilon$-well-supported Nash equilibrium $(\mathbf{x}, \mathbf{y})$, $\mathbf{x}[v_3] = \min(\mathbf{x}[v_1] + \mathbf{x}[v_2], \mathbf{x}_C[v_3]) \pm \epsilon$.*

*Proof.* Properties **1)**–**4)** show that, in any mixed strategy profile $(\mathbf{x}, \mathbf{y})$, we have

$$\langle \mathbf{b}_{2w-1} | \mathbf{x} \rangle - \langle \mathbf{b}_{2w} | \mathbf{x} \rangle = \mathbf{x}[v_1] + \mathbf{x}[v_2] - \mathbf{x}[v_3]$$
$$\langle \mathbf{a}_{2v_3-1} | \mathbf{y} \rangle - \langle \mathbf{a}_{2v_3} | \mathbf{y} \rangle = \mathbf{y}[w] - (\mathbf{y}_C[w] - \mathbf{y}[w])$$

If $\mathbf{x}[v_3] > \min(\mathbf{x}[v_1] + \mathbf{x}[v_2], \mathbf{x}_C[v_3]) + \epsilon$, then $\mathbf{x}[v_3] > \mathbf{x}[v_1] + \mathbf{x}[v_2] + \epsilon$ as $\mathbf{x}[v_3] \leq \mathbf{x}_C[v_3]$. The first equation shows that $\mathbf{y}[w] = 0$ and then the second one implies $\mathbf{x}[v_3] = 0$, according to the definition of well-supported Nash equilibriums. This contradicts with the assumption that $\mathbf{x}[v_3] > \mathbf{x}[v_1] + \mathbf{x}[v_2] + \epsilon > 0$.

If $\mathbf{x}[v_3] < \min(\mathbf{x}[v_1] + \mathbf{x}[v_2], \mathbf{x}_C[v_3]) - \epsilon \leq \mathbf{x}[v_1] + \mathbf{x}[v_2] - \epsilon$, then the first equation shows $\mathbf{y}[w] = \mathbf{y}_C[w]$ and the second one implies $\mathbf{x}[v_3] = \mathbf{x}_C[v_3]$. As $\mathbf{x}_C[v_3] = \mathbf{x}[v_3] < \mathbf{x}[v_1] + \mathbf{x}[v_2]$, we have $\mathbf{x}[v_3] = \mathbf{x}_C[v_3] > \mathbf{x}_C[v_3] - \epsilon = \min(\mathbf{x}[v_1] + \mathbf{x}[v_2], \mathbf{x}_C[v_3]) - \epsilon$, which contradicts with our assumption. □

**Proposition 2 (Gadget $G_\zeta$ where $\zeta \leq 1/K - \epsilon$).** *Let $\mathcal{G}' = (\mathbf{A}, \mathbf{B})$ be a bimatrix game in $\mathcal{L}$ and nodes $v \in V_A$, $w \in V_I$. Let pure strategy profiles $s_1 = (2v - 1, 2w - 1)$, $s_2 = (2v - 1, 2w)$ and $s_3 = (2v, 2w - 1)$. If $\mathcal{G}'$ satisfies*

1). $b_{s_1} = b^*_{s_1} + 1$ and $b_{i,2w-1} = b^*_{i,2w-1}$, for any other $i : 1 \leq i \leq N$;

2). $b_{i,2w} = b^*_{i,2w} + \zeta$, for any $i : 1 \leq i \leq N$;

3). $a_{s_2} = a^*_{s_2} + 1$ and $a_{2v-1,i} = a^*_{2v-1,i}$, for any other $i : 1 \leq i \leq N$;

4). $a_{s_3} = a^*_{s_3} + 1$ and $a_{2v,i} = a^*_{2v,i}$, for any other $i : 1 \leq i \leq N$,

*then in any $\epsilon$-well-supported Nash equilibrium $(\mathbf{x}, \mathbf{y})$, we have $\mathbf{x}[v] = \zeta \pm \epsilon$.*

**Proposition 3 (Gadget $G_{\times \zeta}$, where $\zeta \leq 1$).** *Let $\mathcal{G}' = (\mathbf{A}, \mathbf{B})$ be a bimatrix game in class $\mathcal{L}$ and $v_1, v_2 \in V_A$, $w \in V_I$. Let pure strategy profiles $s_1 = (2v_1 - 1, 2w - 1)$, $s_2 = (2v_2 - 1, 2w)$, $s_3 = (2v_2 - 1, 2w - 1)$ and $s_4 = (2v_2, 2w)$. If $\mathcal{G}'$ satisfies*

1). $b_{s_1} = b^*_{s_1} + \zeta$ and $b_{i,2w-1} = b^*_{i,2w-1}$, for any other $i : 1 \leq i \leq N$;

2). $b_{s_2} = b^*_{s_2} + 1$ and $b_{i,2w} = b^*_{i,2w}$, for any other $i : 1 \leq i \leq N$;

3). $a_{s_3} = a^*_{s_3} + 1$ and $a_{2v_2-1,i} = a^*_{2v_2-1,i}$, for any other $i : 1 \leq i \leq N$;

4). $a_{s_4} = a^*_{s_4} + 1$ and $a_{2v_2,i} = a^*_{2v_2,i}$, for any other $i : 1 \leq i \leq N$,

then in any $\epsilon$-well-supported Nash equilibrium $(\mathbf{x}, \mathbf{y})$, $\mathbf{x}[v_2] = \min(\zeta \mathbf{x}[v_1], \mathbf{x}_C[v_2]) \pm \epsilon$.

**Proposition 4 (Gadget $G_=$).** *Gadget $G_=$ is a special case of gadget $G_{\times \zeta}$. We just set $\zeta$ to be 1, then in any $\epsilon$-well-supported Nash equilibrium $(\mathbf{x}, \mathbf{y})$, $\mathbf{x}[v_2] = \min(\mathbf{x}[v_1], \mathbf{x}_C[v_2]) \pm \epsilon$.*

**Proposition 5 (Gadget $G_-$).** *Let $\mathcal{G}' = (\mathbf{A}, \mathbf{B})$ be a bimatrix game in $\mathcal{L}$ and nodes $v_1, v_2, v_3 \in V_A$, $w \in V_I$. Let pure strategy profiles $s_1 = (2v_1 - 1, 2w - 1)$, $s_2 = (2v_2 - 1, 2w)$, $s_3 = (2v_3 - 1, 2w)$, $s_4 = (2v_3 - 1, 2w - 1)$ and $s_5 = (2v_3, 2w)$. If $\mathcal{G}'$ satisfies*

1). $b_{s_1} = b^*_{s_1} + 1$ and $b_{i,2w-1} = b^*_{i,2w-1}$, for any other $i : 1 \leq i \leq N$;

2). $b_{s_2} = b^*_{s_2} + 1$, $b_{s_3} = b^*_{s_3} + 1$ and $b_{i,2w} = b^*_{i,2w}$, for any other $i : 1 \leq i \leq N$;

3). $a_{s_4} = a^*_{s_4} + 1$ and $a_{2v_3-1,i} = a^*_{2v_3-1,i}$, for any other $i : 1 \leq i \leq N$;

4). $a_{s_5} = a^*_{s_5} + 1$ and $a_{2v_3,i} = a^*_{2v_3,i}$, for any other $i : 1 \leq i \leq N$,

then in any $\epsilon$-well-supported Nash equilibrium $(\mathbf{x}, \mathbf{y})$ of game $\mathcal{G}'$, we have

$$\min(\mathbf{x}[v_1] - \mathbf{x}[v_2], \mathbf{x}_C[v_3]) - \epsilon \leq \mathbf{x}[v_3] \leq \max(\mathbf{x}[v_1] - \mathbf{x}[v_2], 0) + \epsilon.$$

**Proposition 6 (Gadget $G_<$).** *Let $\mathcal{G}' = (\mathbf{A}, \mathbf{B})$ be a bimatrix game in $\mathcal{L}$ and nodes $v_1, v_2, v_3 \in V_A$, $w \in V_I$. Let pure strategy profiles $s_1 = (2v_1 - 1, 2w - 1)$, $s_2 = (2v_2 - 1, 2w)$, $s_3 = (2v_3 - 1, 2w)$ and $s_4 = (2v_3, 2w - 1)$. If $\mathcal{G}'$ satisfies*

1). $b_{s_1} = b^*_{s_1} + 1$ and $b_{i,2w-1} = b^*_{i,2w-1}$, for any other $i : 1 \leq i \leq N$;

2). $b_{s_2} = b^*_{s_2} + 1$ and $b_{i,2w} = b^*_{i,2w}$, for any other $i : 1 \leq i \leq N$;

3). $a_{s_3} = a^*_{s_3} + 1$ and $a_{2v_3-1,i} = a^*_{2v_3-1,i}$, for any other $i : 1 \leq i \leq N$;

4). $a_{s_4} = a^*_{s_4} + 1$ and $a_{2v_3,i} = a^*_{2v_3,i}$, for any other $i : 1 \leq i \leq N$,

then in any $\epsilon$-well-supported Nash equilibrium $(\mathbf{x}, \mathbf{y})$ of game $\mathcal{G}'$, we have $\mathbf{x}[v_3] = \mathbf{x}_C[v_3]$ if $\mathbf{x}[v_1] < \mathbf{x}[v_2] - \epsilon$, and $\mathbf{x}[v_3] = 0$ if $\mathbf{x}[v_1] > \mathbf{x}[v_2] + \epsilon$.

**Proposition 7 (Gadget $G_\vee$).** *Let $\mathcal{G}' = (\mathbf{A}, \mathbf{B})$ be a bimatrix game in $\mathcal{L}$ and nodes $v_1, v_2, v_3 \in V_A$, $w \in V_I$. Let pure strategy profiles $s_1 = (2v_1 - 1, 2w - 1)$, $s_2 = (2v_2 - 1, 2w - 1)$, $s_3 = (2v_3 - 1, 2w - 1)$ and $s_4 = (2v_3, 2w)$. If $\mathcal{G}'$ satisfies*

1). $b_{s_1} = b^*_{s_1} + 1$, $b_{s_2} = b^*_{s_2} + 1$ and $b_{i,2w-1} = b^*_{i,2w-1}$, for any other $i : 1 \leq i \leq N$;

2). $b_{i,2w} = b^*_{i,2w} + 1/(2K)$, for any $i : 1 \leq i \leq N$;

3). $a_{s_3} = a^*_{s_3} + 1$ and $a_{2v_3-1,i} = a^*_{2v_3-1,i}$, for any other $i : 1 \leq i \leq N$;

4). $a_{s_4} = a^*_{s_4} + 1$ and $a_{2v_3,i} = a^*_{2v_3,i}$, for any other $i : 1 \leq i \leq N$,

*then in any $\epsilon$-well-supported Nash equilibrium $(\mathbf{x}, \mathbf{y})$, we have $\mathbf{x}[v_3] = \mathbf{x}_C[v_3]$ if $\mathbf{x}[v_1] = \mathbf{x}_C[v_1]$ or $\mathbf{x}[v_2] = \mathbf{x}_C[v_2]$, and $\mathbf{x}[v_3] = 0$ if $\mathbf{x}[v_1] = \mathbf{x}[v_2] = 0$.*

**Proposition 8 (Gadget $G_\wedge$).** *The construction of $G_\wedge$ is similar as $G_\vee$. We only change the constant in **2)** of Proposition 7 from $1/(2K)$ to $3/(2K)$. In any $\epsilon$-well-supported Nash equilibrium $(\mathbf{x}, \mathbf{y})$, $\mathbf{x}[v_3] = 0$ if $\mathbf{x}[v_1] = 0$ or $\mathbf{x}[v_2] = 0$, and $\mathbf{x}[v_3] = \mathbf{x}_C[v_3]$ if $\mathbf{x}[v_1] = \mathbf{x}_C[v_1]$ and $\mathbf{x}[v_2] = \mathbf{x}_C[v_2]$.*

**Proposition 9 (Gadget $G_\neg$).** *Let $\mathcal{G}' = (\mathbf{A}, \mathbf{B})$ be a game in class $\mathcal{L}$ and $v_1, v_2 \in V_A$, $w \in V_I$. Let pure strategy profiles $s_1 = (2v_1 - 1, 2w - 1)$, $s_2 = (2v_1, 2w)$, $s_3 = (2v_2 - 1, 2w)$ and $s_4 = (2v_2, 2w - 1)$. If $\mathcal{G}'$ satisfies*

1). $b_{s_1} = b^*_{s_1} + 1$ and $b_{i,2w-1} = b^*_{i,2w-1}$, for any other $i : 1 \leq i \leq N$;

2). $b_{s_2} = b^*_{s_2} + 1$ and $b_{i,2w} = b^*_{i,2w}$, for any other $i : 1 \leq i \leq N$;

3). $a_{s_3} = a^*_{s_3} + 1$ and $a_{2v_2-1,i} = a^*_{2v_2-1,i}$, for any other $i : 1 \leq i \leq N$;

4). $a_{s_4} = a^*_{s_4} + 1$ and $a_{2v_2,i} = a^*_{2v_2,i}$, for any other $i : 1 \leq i \leq N$,

*then in any $\epsilon$-well-supported Nash equilibrium $(\mathbf{x}, \mathbf{y})$, $\mathbf{x}[v_2] = 0$ if $\mathbf{x}[v_1] = \mathbf{x}_C[v_1]$ and $\mathbf{x}[v_2] = \mathbf{x}_C[v_2]$ if $\mathbf{x}[v_1] = 0$.*

## 4.5 A Network of Gadgets

In this subsection, we discuss the basic components and methods in building a network of gadgets. Our objective is to encode points in $B^n$ and to simulate the input Boolean circuit $C$ with a bimatrix game $\mathcal{G}$. We will prove a lemma that gives a sufficient condition on the approximate equilibria of $\mathcal{G}$ for a successful encoding and simulation. This lemma enables us to understand the limitations of the gadgets discussed in the last subsection and to work around them in our reduction.

We start with some notations. Let $G_\zeta(v, w)$ denote the insertion of a $G_\zeta$ gadget into $\mathcal{G}^*$ with $v$ as its output node and $w$ as its interior node. For $G_{\times\zeta}$, $G_\neg$ and $G_=$, the gadgets with one input node, let $G(v_1, v_2, w)$ denote the insertion of such a gadget into game $\mathcal{G}^*$ with $v_1$, $v_2$, $w$, respectively, as its input node, output node and interior node. For the rest of gadgets with two input nodes, let $G(v_1, v_2, v_3, w)$ denote the insertion of such a gadget into game $\mathcal{G}^*$ with $v_1$ and $v_2$ as its first and second input node, $v_3$ as its output node and $w$ as its interior node.

Recall $B^n = \mathbb{Z}^n_{[0,7]}$. For an integer lattice point $\mathbf{q} \in B^n$, we will use $\Delta^+_i[\mathbf{q}]$ and $\Delta^-_i[\mathbf{q}]$ to denote the output bits $\Delta^+_i$ and $\Delta^-_i$ of circuit $C$ evaluated at $\mathbf{q}$.

For a probability vector $\mathbf{x} \in \mathbb{P}^N$ and a Boolean $b$, we use $\mathbf{x}[v] =_B b$ to denote $\mathbf{x}[v] = \mathbf{x}_C[v]$ if $b$ is boolean 1 and $\mathbf{x}[v] = 0$ if $b$ is boolean 0.

**Definition 11 (Well-Positioned Points).** *A real number $a \in \mathbb{R}^+$ is poorly-positioned if there exists an integer $0 \leq t \leq 7$ such that $|a - t| \leq 80K\epsilon = 80/K^2$. A point $\mathbf{p} \in \mathbb{R}^n_+$ is well-positioned if none of its components is poorly-positioned. If $\mathbf{p}$ is not well-positioned, we refer to it as a poorly-positioned point.*

For each $a \in \mathbb{R}^+$, let $\pi(a)$ be the largest integer in $[0:7]$ that is smaller than $a$, that is

$$\pi(a) = \max\{\, i \mid 0 \leq i \leq 7 \text{ and } i < a \,\}.$$

As we have discussed earlier, we use each node of $V_A$ as an output in at most one gadget and use each node of $V_I$ in at most one gadget. In addition, the insertion of a gadget only modifies the two rows and the two columns associated with its output node and its internal node. Because the notion of $\epsilon$-well-supported Nash equilibria is defined by a pairwise condition, it allows us to argue about the structure of an $\epsilon$-well-supported Nash equilibrium locally, even without the full knowledge of the rest of the game. So, consider an $\epsilon$-well-supported Nash equilibrium $(\mathbf{x}, \mathbf{y})$ of our ultimate game. Suppose $v, v_1, v_2, v_3$ are four arithmetic nodes in $V_A$. Let $a = 8K\mathbf{x}[v]$. By lemma 6, $0 \leq a \leq 8K(\frac{1}{K} + \epsilon) = 8 + 1/K^2$. Figure 8 presents a small network to compute $\pi(a)$ with the help of some additional nodes in the network. The following lemma states that the values of $v_1, v_2, v_3$ are exactly the binary representation of $\pi(a)$ as long as $a$ is not poorly-positioned.

**Lemma 7 (Encoding Binary with Games).** *In any $\epsilon$-well-supported Nash equilibrium $(\mathbf{x}, \mathbf{y})$, if $a = 8K\mathbf{x}[v] \in \mathbb{R}^+$ is not poorly-positioned, then we have $\mathbf{x}[v_i] =_B b_i$, where $b_1 b_2 b_3$ is the binary representation of integer $\pi(a)$.*

*Proof.* First, we consider the case when $\pi(a) = 7$. As $a \geq 7 + 80K\epsilon$, we have $\mathbf{x}[v] \geq 1/(2K) + 1/(4K) + 1/(8K) + 10\epsilon$. From figure 8, we have $\mathbf{x}[v_1^*] \geq \mathbf{x}[v] - \epsilon$, $\mathbf{x}[v_1] =_B 1$ in the first loop and

$$\begin{aligned}
\mathbf{x}[v_2^*] &\geq \mathbf{x}[v_1^*] - \mathbf{x}[v_1^2] - \epsilon \geq \mathbf{x}[v] - \epsilon - (2^{-1}\mathbf{x}[v_1] + \epsilon) - \epsilon \\
&\geq \mathbf{x}[v] - 2^{-1}(1/K + \epsilon) - 3\epsilon > 1/(4K) + 1/(8K) + 6\epsilon.
\end{aligned}$$

Since $\mathbf{x}[v_2^1] \leq 1/(4K) + \epsilon$ and $\mathbf{x}[v_2^*] - \mathbf{x}[v_2^1] > \epsilon$, we have $\mathbf{x}[v_2] =_B 1$ and

$$\mathbf{x}[v_3^*] \geq \mathbf{x}[v_2^*] - \mathbf{x}[v_2^2] - \epsilon > 1/(8K) + 3\epsilon.$$

As a result, $\mathbf{x}[v_3^*] - \mathbf{x}[v_3^1] > \epsilon$ and $\mathbf{x}[v_3] =_B 1$.

Next, we consider the general case that $t < \pi(a) < t + 1$ for $0 \leq t \leq 6$. Let $b_1 b_2 b_3$ be the binary representation of $t$. As $a$ is well-positioned, we have

$$b_1/(2K) + b_2/(4K) + b_3/(8K) + 10\epsilon \leq \mathbf{x}[v] \leq b_1/(2K) + b_2/(4K) + (b_3 + 1)/(8K) - 10\epsilon.$$

With similar arguments, after the first loop, one can show that $\mathbf{x}[v_1] =_B b_1$ and

$$b_2/(4K) + b_3/(8K) + 6\epsilon \leq \mathbf{x}[v] \leq b_2/(4K) + (b_3 + 1)/(8K) - 6\epsilon.$$

After the second loop, we have $\mathbf{x}[v_2] =_B b_2$ and

$$b_3/(8K) + 3\epsilon \leq \mathbf{x}[v] \leq (b_3 + 1)/(8K) - 3\epsilon.$$

Thus, $\mathbf{x}[v_3] =_B b_3$. $\square$

**A Network Which Computes $\pi(a)$**

---

1: pick unused nodes $v_1^*, v_2^*, v_3^*, v_4^* \in V_A$ and $w \in V_I$
2: insert gadget $G_=(v, v_1^*, w)$
3: **for** $j$ from 1 to 3 **do**
4:    pick unused nodes $v^1, v^2 \in V_A$ and $w^1, w^2, w^3, w^4 \in V_I$
5:    insert gadgets $G_{2^{-(6m+i)}}(v^1, w^1)$, $G_<(v^1, v_i^*, v_i, w^2)$
6:    insert gadgets $G_{\times 2^{-i}}(v_i, v^2, w^3)$, $G_-(v_i^*, v^2, v_{i+1}^*, w^4)$

---

Figure 8: A Network Which Computes $\pi(a)$

Note that because the comparator $G_<$ is brittle, the values of $v_i$ could be arbitrary if $a$ is poorly-positioned.

For a well-positioned point $\mathbf{p} \in \mathbb{R}_+^n$, let $\mathbf{q} = \pi(\mathbf{p})$ be the integer lattice point in $B^n$ with $q_i = \pi(p_i)$ for all $i : 1 \leq i \leq n$. We now construct a larger network of gadgets to simulate the evaluation of circuit $C$. Let $\{v_i\}_{1 \leq i \leq n}$ and $\{v_i^+, v_i^-\}_{1 \leq i \leq n}$ be $3n$ arithmetic nodes. For a pair of probability vectors $(\mathbf{x}, \mathbf{y})$, we view the values of $\{v_i\}_{1 \leq i \leq n}$ as an encoding of a point $\mathbf{p} \in \mathbb{R}_+^n$, with $p_i = 8K\mathbf{x}[v_i]$. This network then guarantees that in a $\epsilon$-well-supported Nash equilibrium $(\mathbf{x}, \mathbf{y})$ of the ultimate game, if $\mathbf{p}$ is a well-positioned point, then

$$\mathbf{x}[v_i^+] =_B \Delta_i^+[\mathbf{q}] \quad \text{and} \quad \mathbf{x}[v_i^-] =_B \Delta_i^-[\mathbf{q}], \quad \text{where } \mathbf{q} = \pi(\mathbf{p}) \in B^n. \tag{3}$$

The network is divided into two parts.

**Part 1.** Let $\{v_{i,j}\}_{1 \leq i \leq n,\ 1 \leq i \leq 3}$ to be $3n$ arithmetic nodes. For each $k : 1 \leq k \leq n$, we add a network described in Figure 8 with connect $v_k$ with $v_{k,1}, v_{k,2}, v_{k,3}$ so that the values at $v_{k,1}, v_{k,2}, v_{k,3}$ encode the binary bits of the number represented by $v_k$ according to Lemma 7.

**Part 2.** We view the values of the $3n$ nodes $\{v_{i,j}\}_{1 \leq i \leq n,\ 1 \leq j \leq 3}$ as the encoding of $3n$ input bits of circuit $C$, and use the three type of logic gadgets $G_\wedge, G_\vee, G_\neg$ to simulate the evaluation of $C$ on these bits. The $2n$ output bits are stored in arithmetic nodes $\{v_i^+, v_i^-\}_{1 \leq i \leq n}$. The simulation of $C$ works correctly in any $\epsilon$-well-supported Nash equilibrium, provided that the values of nodes $\{v_{i,j}\}_{1 \leq i \leq n,\ 1 \leq j \leq 3}$ are representations of boolean bits.

Equation 3 is a direct corollary of Lemma 7. We refer to this simulation network as a *sampling network*. Note that this network works correctly only if $\mathbf{p}$ is a well-positioned point, and when $\mathbf{p}$ is not, the values of $\{v_i^+, v_i^-\}_{1 \leq i \leq n}$ could be arbitrary. In which case, we only know, or at least fortunately know, that $0 \leq \mathbf{x}[v_i^+], \mathbf{x}[v_i^-] \leq 1/K + \epsilon$, because the game we design belongs to class $\mathcal{L}$.

## 4.6 Construction of the Game $\mathcal{G}$

We place $n^4$ distinguished arithmetic nodes $\{v_i^k\}_{0\le k<n^3,\ 1\le i\le n}$ in the game $\mathcal{G}$. In an $\epsilon$-well-supported Nash equilibrium $(\mathbf{x},\mathbf{y})$ of $\mathcal{G}$, they encode $n^3$ points $S=\{\mathbf{p}^k\}_{0\le k<n^3}$, where $p_i^k = 8K\mathbf{x}[v_i^k]$ for all $i: 1\le i\le n$. Our objective is to design the game $\mathcal{G}$ so that it satisfies the following property.

**Property 3.** *Let $(\mathbf{x},\mathbf{y})$ be an $\epsilon$-well-supported Nash equilibrium of $\mathcal{G}$. Then*

$$Q = \left\{\, \mathbf{q}^k = \pi(\mathbf{p}^k) \,\bigg|\, \mathbf{p}^k \text{ is a well-positioned point},\ 0\le k < n^3 \,\right\}$$

*is a panchromatic set of $C$*

*Proof.* It follows directly from Lemmas 12, 11, and 13. □

We introduce some more notations for our construction below. For two vectors $\mathbf{r},\mathbf{r}'\in\mathbb{R}^n$ and $a>0$, we will use $\mathbf{r}=\mathbf{r}'\pm a$ to denote $r_i = r_i'\pm a$ for all $1\le i\le n$.

Let $I_G$ and $I_B$ denote, respectively, the sets of indices of well-positioned and poorly-positioned points in $S$, that is,

$$\begin{aligned} I_G &= \left\{\, t \,\bigg|\, 0\le t < n^3,\ \mathbf{p}^t\in S \text{ is a well-positioned point}\,\right\}, \quad\text{and}\\ I_B &= \left\{\, t \,\bigg|\, 0\le t < n^3,\ \mathbf{p}^t\in S \text{ is a poorly-positioned point}\,\right\}. \end{aligned}$$

For each $t\in I_G$, let $c_t\in\{1,2,...,n,n+1\}$ be the color of $\mathbf{q}^t = \pi(\mathbf{p}^t)\in B^n$ assigned by circuit $C$. We also define, for each $i: 1\le i\le n+1$, $T_i = |\{t\in I_G\,|\,c_t = i\}|$ to be the number of well-positioned points in $S$ whose associated integer lattice point in $B^n$ is colored with $i$.

Our construction of $\mathcal{G}$ is divided into the following four parts.

**Part 1.** For each $0 < k < n^3$ and $1\le i\le n$, by inserting gadgets $G_\zeta$ and $G_+$, we make sure

$$\mathbf{x}[v_i^k] = \min\left(\mathbf{x}[v_i^0] + \frac{k}{K^2}, \mathbf{x}_C[v_i^k]\right) \pm O(\epsilon). \tag{4}$$

in any $\epsilon$-well-supported Nash equilibrium $(\mathbf{x},\mathbf{y})$ of $\mathcal{G}$. We use $1/K^2$ as the increment in this step because it is much larger than $\epsilon = 1/K^3$ and much smaller than $1/(8K)$. This property of our choice of increment is very important in proving the following two lemmas.

**Lemma 8 (Not Too Many Poorly-Positioned Points).** *In any $\epsilon$-well-supported Nash equilibrium $(\mathbf{x},\mathbf{y})$ of $\mathcal{G}$, $|I_B|\le n$.*

*Proof.* For each $t\in I_B$, according to the definition of poorly-positioned point, there exists an integer $1\le l\le n$ such that $p_l^t$ is a poorly-positioned number. We will prove that, for every integer $1\le l\le n$, there exists at most one $0\le t < n^3$ such that real number $p_l^t = 8K\mathbf{x}[v_l^t]$ is poorly-positioned, which implies $|I_B|\le n$ immediately.

Assume $p_l^t$ and $p_l^{t'}$ are both poorly-positioned, for a pair of integers $0\le t < t' < n^3$. Then, from the definition, there exists a pair of integers $0\le k,k'\le 7$,

$$\left|\mathbf{x}[v_l^t] - k/(8K)\right| \le 10\epsilon \quad\text{and}\quad \left|\mathbf{x}[v_l^{t'}] - k'/(8K)\right|\le 10\epsilon. \tag{5}$$

Because (5) implies that $\mathbf{x}[v_l^t] < 1/K - \epsilon \leq \mathbf{x}_C[v_l^t]$ and $\mathbf{x}[v_l^{t'}] < 1/K - \epsilon \leq \mathbf{x}_C[v_l^{t'}]$, by Equation (4) of **Part 1**, we have

$$\mathbf{x}[v_l^t] = \mathbf{x}[v_l^0] + t/K^2 \pm O(\epsilon) \quad \text{and} \quad \mathbf{x}[v_l^{t'}] = \mathbf{x}[v_l^0] + t'/K^2 \pm O(\epsilon). \tag{6}$$

Hence, $\mathbf{x}[v_l^t] < \mathbf{x}[v_l^{t'}]$, $k \leq k'$ and

$$\mathbf{x}[v_l^{t'}] - \mathbf{x}[v_l^t] = (t' - t)/K^2 \pm O(\epsilon) \tag{7}$$

Note that when $k = k'$, Equation (5) implies that $\mathbf{x}[v_l^{t'}] - \mathbf{x}[v_l^t] \leq 20\epsilon$, while when $k < k'$, it implies that $\mathbf{x}[v_l^{t'}] - \mathbf{x}[v_l^t] \geq (k'-k)/(8K) - 20\epsilon \geq 1/(8K) - 20\epsilon$. In both cases, we derived an inequality that contradicts with (7). Thus, only one of $p_l^t$ or $p_l^{t'}$ can be poorly-positioned. □

**Lemma 9** ($Q$ **Is Accommodated**). *In any $\epsilon$-well-supported Nash equilibrium, $Q$ is accommodated and $|Q| \leq n+1$.*

*Proof.* To show $Q$ is accommodated, it is sufficient to prove

$$q_l^t \leq q_l^{t'} \leq q_l^t + 1, \qquad \text{for all } 1 \leq l \leq n \text{ and } t < t' \in I_G. \tag{8}$$

First, assume $q_l^t > q_l^{t'}$ for some pair of $t < t'$ and $t, t' \in I_G$. Since $q_l^{t'} < q_l^t \leq 7$, we have $p_l^{t'} < 7$ and thus, $\mathbf{x}[v_l^{t'}] < 7/(8K)$. As a result, the first component of the min operator in (4) is the smallest for both $t$ and $t'$, implying that $\mathbf{x}[v_l^t] < \mathbf{x}[v_l^{t'}]$ and $p_l^t < p_l^{t'}$, which results in a contradiction with the assumption that $q_l^t > q_l^{t'}$.

Second, assume $q_l^{t'} - q_l^t \geq 2$ for some pair of $t < t'$ and $t, t' \in I_G$. From to the definition of $\pi$, we have $p_l^{t'} - p_l^t > 1$ and thus, $\mathbf{x}[v_l^{t'}] - \mathbf{x}[v_l^t] > 1/(8K)$. But from (4), we have $\mathbf{x}[v_l^{t'}] - \mathbf{x}[v_l^t] \leq (t'-t)/K^2 + O(\epsilon) < n^3/K^2 + O(\epsilon) \ll 1/(8K)$. Thus, (8) is true.

Now we prove $|Q| \leq n+1$. Note that the definition of $Q$ together with (4) implies that there exists integers $t_1 < t_2 < ... < t_{|Q|} \in I_G$ such that $\mathbf{q}^{t_i}$ is strictly dominated by $\mathbf{q}^{t_{i+1}}$, that is, $\mathbf{q}^{t_i} \neq \mathbf{q}^{t_{i+1}}$ and $q_j^{t_i} \leq q_j^{t_{i+1}}$ for all $j : 1 \leq j \leq n$.

On the one hand, for every $1 \leq l \leq |Q| - 1$, there exists an integer $1 \leq k_j \leq n$ such that $q_{k_j}^{t_{l+1}} = q_{k_j}^{t_l} + 1$. On the other hand, for every $1 \leq k \leq n$, (4) implies that there is at most one $1 \leq l \leq |Q| - 1$ such that $q_k^{t_{l+1}} = q_k^{t_l} + 1$. Therefore, $|Q| \leq n+1$. □

By Lemma 9, to prove Property 3, it is sufficient to establish in any $\epsilon$-well-supported Nash equilibrium $(\mathbf{x}, \mathbf{y})$, $T_i > 0$ for all $1 \leq i \leq n+1$.

**Part 2.** We allocate $2n^4$ unused arithmetic nodes $\{v_i^{k+}, v_i^{k-}\}_{1 \leq i \leq n,\ 0 \leq k < n^3}$. For each integer $k : 0 \leq k < n^3$, we insert a sampling network (see Section 4.5) to connect $\{v_i^k\}_{1 \leq i \leq n}$ with $\{v_i^{k+}, v_i^{k-}\}_{1 \leq i \leq n}$ in order to simulate the evaluation of $C$ on the point associated with $\{v_i^k\}_{1 \leq i \leq n}$.

For any $0 \leq k < n^3$, we use $\mathbf{r}^k$ to denote the vector that satisfies $r_i^k = \mathbf{x}[v_i^{k+}] - \mathbf{x}[v_i^{k-}]$ for all $i : 1 \leq i \leq n$. Let $E^n = \{\mathbf{z}^1, \mathbf{z}^2, ..., \mathbf{z}^n, \mathbf{z}^{n+1}\}$ be a set of $n+1$ vectors where $\mathbf{z}^i = \mathbf{e}_i/K$ and $\mathbf{z}^{n+1} = -\sum_{1 \leq i \leq n} \mathbf{e}_i/K$. The next lemma follows directly from Section 4.5 and the definition of valid Brouwer-mapping circuits.

**Lemma 10 (Correct Encoding of Colors).** *Let $(\mathbf{x}, \mathbf{y})$ be an $\epsilon$-well-supported Nash equilibrium. Then for any $t \in I_G$, $\mathbf{r}^t = \mathbf{z}^{c_t} \pm \epsilon$.*

In the next step, we add all these vectors from Part 2 together in order to reduce the contribution from poorly-positioned points.

**Part 3.** Let $\{v_i^+, v_i^-\}_{1 \leq i \leq n}$ be $2n$ unused arithmetic nodes. By inserting gadgets $G_{\times \zeta}$ and $G_+$, we make sure

$$\mathbf{x}[v_i^+] = \sum_{0 \leq k < n^3} \left(\frac{1}{K} \mathbf{x}[v_i^{k+}]\right) \pm O(n^3 \epsilon) \quad \text{and} \quad \mathbf{x}[v_i^-] = \sum_{0 \leq k < n^3} \left(\frac{1}{K} \mathbf{x}[v_i^{k-}]\right) \pm O(n^3 \epsilon)$$

in any $\epsilon$-well-supported Nash equilibrium $(\mathbf{x}, \mathbf{y})$. Note that the multiplication gadget $G_{\times 1/K}$ should be inserted before $G_+$.

Finally, let $\mathbf{r}$ denote the vector with $r_i = \mathbf{x}[v_i^+] - \mathbf{x}[v_i^-]$, for all $i : 1 \leq i \leq n$. In the last part of our construction, we insert comparison gadgets (together addition and subtraction gadgets) to ensure $\|\mathbf{r}\|_\infty = \max_i |r_i|$ is close to zero in any $\epsilon$-well-supported Nash equilibrium $(\mathbf{x}, \mathbf{y})$ of $\mathcal{G}$.

**Part 4.** For each $1 \leq i \leq n$, we pick two unused nodes $v_i', v_i'' \in V_A$ and $w_1, w_2, w_3 \in V_I$, and insert the following three gadgets:

$$G_+(v_i^0, v_i^+, v_i', w_1), \qquad G_-(v_i', v_i^-, v_i'', w_2), \qquad G_=(v_i'', v_i^0, w_3).$$

Ideally, we wish to establish $\|\mathbf{r}\|_\infty = O(\epsilon)$ as one might hope. However, whether or not this condition holds depends on the value of nodes $v_i^0$. For example, in the case when $\mathbf{x}[v_i^0] = 0$, the magnitude of $\mathbf{x}[v_i^-]$ could be much larger than that of $\mathbf{x}[v_i^+]$. We are able to establish the following lemma which is sufficient to carry out our correctness proof of the reduction.

**Lemma 11 (Well-Conditioned Solution).** *For an $\epsilon$-well-supported Nash equilibrium $(\mathbf{x}, \mathbf{y})$ of $\mathcal{G}$ and for all $i : 1 \leq i \leq n$,*

1. *if $\mathbf{x}[v_i^0] > 4\epsilon$, then $r_i = \mathbf{x}[v_i^+] - \mathbf{x}[v_i^-] > -4\epsilon$;*

2. *if $\mathbf{x}[v_i^0] < 1/K - 2n^3/K^2$, then $r_i = \mathbf{x}[v_i^+] - \mathbf{x}[v_i^-] < 4\epsilon$.*

*Proof.* In order to set up a proof-by-contradiction of the first if-statement of this lemma, we assume there exists some $i$ such that $\mathbf{x}[v_i^0] > 4\epsilon$ and $\mathbf{x}[v_i^+] - \mathbf{x}[v_i^-] \leq -4\epsilon$.

By the first gadget $G_+(v_i^0, v_i^+, v_i', w_1)$, we have

$$\mathbf{x}[v_i'] = \min(\mathbf{x}[v_i^0] + \mathbf{x}[v_i^+], \mathbf{x}_C[v_i']) \pm \epsilon \leq \mathbf{x}[v_i^0] + \mathbf{x}[v_i^+] + \epsilon \leq \mathbf{x}[v_i^0] + \mathbf{x}[v_i^-] - 3\epsilon. \tag{9}$$

By the second gadget $G_-(v_i', v_i^-, v_i'', w_2)$, we have

$$\mathbf{x}[v_i''] \leq \max(\mathbf{x}[v_i'] - \mathbf{x}[v_i^-], 0) + \epsilon \leq \max(\mathbf{x}[v_i^0] - 3\epsilon, 0) + \epsilon \leq \mathbf{x}[v_i^0] - 2\epsilon, \tag{10}$$

where the last inequality follows from the assumption $\mathbf{x}[v_i^0] > 4\epsilon$. So by the upper bound of Lemma 6, $\mathbf{x}[v_i''] \leq \mathbf{x}[v_i^0] - 2\epsilon \leq 1/K - \epsilon \leq \mathbf{x}_C[v_i^0]$. Thus, $\min(\mathbf{x}[v_i''], \mathbf{x}_C[v_i^0]) = \mathbf{x}[v_i'']$.

So, by the last gadget $G_=(v_i'', v_i^0, w_3)$, we have $\mathbf{x}[v_i^0] = \min(\mathbf{x}[v_i''], \mathbf{x}_C[v_i^0]) \pm \epsilon = \mathbf{x}[v_i''] \pm \epsilon$, which contradicts with (10).

Similarly, to prove the second if-statement of the lemma, we assume there exists some $i$ such that $\mathbf{x}[v_i^0] < 1/K - 2n^3/K^2$ and $\mathbf{x}[v_i^+] - \mathbf{x}[v_u^-] \geq 4\epsilon$ in order to derive a contradiction.

By Part 3 and Lemma 6, $\mathbf{x}[v_i^+] \leq n^3/K^2 + O(n^3\epsilon)$. Together with the assumption, we have $\mathbf{x}[v_i^0] + \mathbf{x}[v_i^+] \leq 1/K - n^3/K^2 + O(n^3\epsilon) < 1/K - \epsilon \leq \mathbf{x}_C[v_i']$. Thus, by the first gadget, we have

$$\mathbf{x}[v_i'] = \min(\mathbf{x}[v_i^0] + \mathbf{x}[v_i^+], \mathbf{x}_C[v_i']) \pm \epsilon = \mathbf{x}[v_i^0] + \mathbf{x}[v_i^+] \pm \epsilon \geq \mathbf{x}[v_i^0] + \mathbf{x}[v_i^-] + 3\epsilon$$

and $\mathbf{x}[v_i'] \leq \mathbf{x}[v_i^0] + \mathbf{x}[v_i^+] + \epsilon \leq 1/K - n^3/k^2 + O(n^3\epsilon) < \mathbf{x}_C[v_i'']$. By the second gadget $G_-$, we have

$$\mathbf{x}[v_i''] \geq \min(\mathbf{x}[v_i'] - \mathbf{x}[v_i^-], \mathbf{x}_C[v_i'']) - \epsilon = \mathbf{x}[v_i'] - \mathbf{x}[v_i^-] - \epsilon \geq \mathbf{x}[v_i^0] + 2\epsilon, \tag{11}$$

since $\mathbf{x}[v_i'] - \mathbf{x}[v_i^-] \leq \mathbf{x}[v_i'] < \mathbf{x}_C[v_i'']$. But the last gadget $G_=$ implies

$$\mathbf{x}[v_i^0] = \min(\mathbf{x}[v_i''], \mathbf{x}_C[v_i^0]) \pm \epsilon = \mathbf{x}[v_i''] \pm \epsilon,$$

which contradicts with (11), where $\min(\mathbf{x}[v_i''], \mathbf{x}_C[v_i^0]) = \mathbf{x}[v_i'']$ because from the second gadget $G_-$, we have $\mathbf{x}[v_i''] \leq \max(\mathbf{x}[v_i'] - \mathbf{x}[v_i^-], 0) + \epsilon \leq \mathbf{x}[v_i'] + \epsilon \leq 1/K - n^3/k^2 + O(n^3\epsilon) + \epsilon < \mathbf{x}_C[v_i^0]$. □

### 4.7 Correctness of the Reduction

First, adding up the number of nodes of $V_A$ and hence $V_I$ in our construction, we see that at most $K$ arithmetic nodes and $K$ internal nodes are used in our construction. Thus, size of the bimatrix game $\mathcal{G}$ is at most $N = 2K$ which is polynomial in $\text{Size}[C]$. Moreover, $\mathcal{G}$ can be constructed from the input instance $(C, 0^n)$ of BROUWER in time polynomial in $\text{Size}[C]$. Thus., in order to prove the correctness of our reduction, it is sufficient to establish Property 3.

For any $\epsilon$-well-supported Nash equilibrium $(\mathbf{x}, \mathbf{y})$, by **Part 3** and **Part 4** of our construction, we have

$$\begin{aligned}
\mathbf{r} &= \frac{1}{K}\left(\sum_{0 \leq i < n^3} \mathbf{r}^i\right) \pm O(n^3\epsilon) = \frac{1}{K}\left(\sum_{i \in I_G} \mathbf{r}^i\right) + \frac{1}{K}\left(\sum_{i \in I_B} \mathbf{r}^i\right) \pm O(n^3\epsilon) \\
&= \frac{1}{K}\left(\sum_{i \in I_G} \mathbf{z}^{c_i}\right) + \frac{1}{K}\left(\sum_{i \in I_B} \mathbf{r}^i\right) \pm O(n^3\epsilon) \\
&= \frac{1}{K}\left(\sum_{1 \leq i \leq n+1} T_i \mathbf{z}^i\right) + \frac{1}{K}\left(\sum_{i \in I_B} \mathbf{r}^i\right) \pm O(n^3\epsilon). \tag{12}
\end{aligned}$$

Let $\mathbf{r}^G = \frac{1}{K}\left(\sum_{1 \leq i \leq n+1} T_i \mathbf{z}^i\right)$ and $\mathbf{r}^B = \frac{1}{K}\left(\sum_{i \in I_B} \mathbf{r}^i\right)$. So we have $\mathbf{r} = \mathbf{r}^G + \mathbf{r}^B \pm O(n^3\epsilon)$. Since $|I_B| \leq n$ and $\|\mathbf{r}^i\|_\infty \leq 1/K + \epsilon$ according to lemma 6, we obtain $\|\mathbf{r}^B\|_\infty = O(n/K^2)$.

Because $|I_G| \geq n^3 - |I_B| \geq n^3 - n$, we have $\sum_{1 \leq i \leq n+1} T_i \geq n^3 - n$. The next lemma shows that, if one of $T_i$ equals zero, then $\|\mathbf{r}^G\|_\infty \gg \|\mathbf{r}^B\|_\infty$.

**Lemma 12 (Color Gap).** *Let* $\mathbf{r}' = \sum_{1 \leq i \leq n+1} T_i \mathbf{z}^i$. *If one of $T_i$ is zero, then* $\|\mathbf{r}'\|_\infty \geq n^2/(3K)$, *and thus* $\|\mathbf{r}^G\|_\infty \geq n^2/(3K^2)$ *and* $\|\mathbf{r}\|_\infty \gg 4\epsilon$.

*Proof.* First, assume $T_{n+1} = 0$. Let $l$ be the integer such that $T_l = \max_{1 \leq i \leq n} T_i$, then $T_l > n^2 - 1$. Thus, $r'_l = T_l/K \geq (n^2-1)/K > n^2/(3K^2)$.

Second, assume $T_t = 0$ for some $1 \leq t \leq n$. We have the following two cases: (1) When $T_{n+1} \geq n^2/2$, $r'_t = -T_{n+1}/K \leq -n^2/(2K^2) < -n^2/(3K^2)$. (2) When $T_{n+1} < n^2/2$. Let $l$ be the integer such that $T_l = \max_{1 \leq i \leq n+1} T_i$, then $l \neq t, n+1$ and $T_l > n^2 - 1$. Then, $r'_l = (T_l - T_{n+1})/K > (n^2/2 - 1)/K^2 > n^2/(3K^2)$. $\square$

Therefore, if $Q$ is not a panchromatic set, then one of the $T_i$'s is equal to zero, and hence $\|\mathbf{r}\|_\infty \geq n^2/(2K^2) \gg \epsilon$. Had the **Part 4** of our construction guaranteed $\|\mathbf{r}\|_\infty = O(\epsilon)$, we would have completed the proof. As it is not always the case, we now prove the following lemma to establish a condition so that we could use Lemma 11 to complete the proof.

**Lemma 13 (Well-Conditioness).** *Let $(\mathbf{x}, \mathbf{y})$ be an $\epsilon$-well-supported Nash equilibrium for $\mathcal{G}$. Then $4\epsilon < \mathbf{x}[v_i^0] < 1/K - 2n^3/K^2$, for all integers $i : 1 \leq i \leq n$.*

*Proof.* In this proof, we will use Lemma 14 below about the boundary conditions of $C$. We will also use $1/K = 2^{-6m}$, $\epsilon = 2^{-18m}$, and $2^m > n$.

First, if there exists an integer $k : 1 \leq k \leq n$ such that $\mathbf{x}[v_k^0] \leq 4\epsilon$, then $q_k^t = 0$ for all $t \in I_G$, according to **Part 1**. By Lemma 14.1, $T_{n+1} = 0$. Let $1 \leq l \leq n$ be the integer such that $T_l = \max_{1 \leq i \leq n} T_i$. As $\sum_{i=1}^{n+1} T_i = n^3$, we have $T_l > n^2 - 1$. So, by Equation (12), $r_l = T_l/K^2 - n/K^2 \pm O(n^3\epsilon) > 4\epsilon$. Consider the following cases:

- If $\mathbf{x}[v_l^0] < 1/K - 2n^3/K^2$, then we get a contradiction in Lemma 11.2.

- If $\mathbf{x}[v_l^0] \geq 1/K - 2n^3/K^2$, then for all $t \in I_G$,

$$p_l^t = 8K\mathbf{x}[v_l^t] = 8K\left(\min(\mathbf{x}[v_l^0] + t/K^2, \mathbf{x}_C[v_l^t]) \pm O(\epsilon)\right) > 1$$

  and hence $q_l^t > 0$. By Lemma 14.2, we have $T_l = 0$ which contradicts with the assumption.

Second, if there exists an integer $k : 1 \leq k \leq n$ such that $\mathbf{x}[v_k^0] \geq 1/K - 2n^3/K^2$, then we have $q_k^t = 7$ for all $t \in I_G$. By Lemma 14.3, $T_k = 0$. If $T_{n+1} \geq n^2/2$, then $r_k = -T_{n+1}/K^2 + n/K^2 \pm O(n^3\epsilon) < -4\epsilon$, which contradicts with the assumption that $\mathbf{x}[v_k^0] \geq 1/K - 2n^3/K^2 > 4\epsilon$ (see Lemma 11.1). Below, we assume $T_{n+1} < n^2/2$.

Let $l$ be the integer such that $T_l = \max_{1 \leq i \leq n+1} T_i$. Since $T_k = 0$, we have $T_l \geq n^2 - 1$ and $l \neq k$. As $T_{n+1} < n^2/2$, $T_l - T_{n+1} > n^2/2 - 1$ and $r_l = (T_l - T_{n+1})/K^2 - n/K^2 \pm O(n^3\epsilon) > 4\epsilon$. Consider the following two cases:

- $\mathbf{x}[v_l^0] < 1/K - 2n^3/K^2$, then we get a contradiction in Lemma 11.2.

- $\mathbf{x}[v_l^0] \geq 1/K - 2n^3/K^2$, then $p_l^t > 1$ and thus $q_l^t > 0$, for all $t \in I_G$. By Lemma 14.4, we have $T_l = 0$ which contradicts with the assumption.

In conclusion, $4\epsilon < \mathbf{x}[v_i^0] < 1/K - 2n^3/K^2$ for all $1 \leq i \leq n$, and the lemma is proven. $\square$

**Lemma 14 (Boundary Conditions).** *For every lattice point $\mathbf{q} \in B^n$ and integers $1 \leq k \neq l \leq n$,*

1. if $q_k = 0$, then $Color_C[\mathbf{q}] \neq n+1$;

2. if $q_k = 0$ and $q_l > 0$, then $Color_C[\mathbf{q}] \neq l$;

3. if $q_k = 7$, then $Color_C[\mathbf{q}] \neq k$;

4. if $q_k = 7$ and $Color_C[\mathbf{q}] = l \neq k$, then $q_l = 0$.

*Proof.* The lemma follows directly from the definition of valid circuits. □

## 5 Smoothed Complexity of Bimatrix Games

In the smoothed analysis of the bimatrix game, we assume that each entry of the payoff matrices is subject to a small and independent random perturbation. Consider a bimatrix game given by two $n \times n$ matrices $(\bar{\mathbf{A}}, \bar{\mathbf{B}})$, where $\bar{\mathbf{A}}, \bar{\mathbf{B}} \in \mathbb{R}_{[-1,1]}^{n \times n}$. In the smoothed model, the input[8] is then defined by $(\mathbf{A}, \mathbf{B})$, where $a_{i,j}$ and $b_{i,j}$ are, respectively, independent perturbations of $\bar{a}_{i,j}$ and $\bar{b}_{i,j}$, with magnitude $\sigma$.

### 5.1 Models of Perturbations and Smoothed Complexity

There might be several models of perturbations for $a_{i,j}$ and $b_{i,j}$ with magnitude $\sigma$. The common two perturbation models are the uniform perturbation and Gaussian perturbation.

In the *uniform perturbation* with magnitude $\sigma$, $a_{i,j}$ and $b_{i,j}$ are chosen uniformly from the intervals $[\bar{a}_{i,j} - \sigma, \bar{a}_{i,j} + \sigma]$ and $[\bar{b}_{i,j} - \sigma, \bar{b}_{i,j} + \sigma]$, respectively. In the *Gaussian perturbation* with variance $\sigma^2$, $a_{i,j}$ and $b_{i,j}$ are, respectively, chosen with density

$$\frac{1}{\sqrt{2\pi}\sigma}e^{-|a_{i,j}-\bar{a}_{i,j}|^2/2\sigma^2} \quad \text{and} \quad \frac{1}{\sqrt{2\pi}\sigma}e^{-|b_{i,j}-\bar{b}_{i,j}|^2/2\sigma^2}.$$

Following Spielman and Teng [23], the smoothed complexity of an algorithm $J$ for the bimatrix game is defined as following: Let $T_J(\mathbf{A}, \mathbf{B})$ be the complexity of algorithm $J$ for solving a bimatrix game defined by $(\mathbf{A}, \mathbf{B})$. Then, the smoothed complexity of algorithm $J$ under perturbations $N_\sigma()$ of magnitude $\sigma$ is

$$\text{Smoothed}_J[n, \sigma] = \max_{\bar{\mathbf{A}}, \bar{\mathbf{B}} \in \mathbb{R}_{[-1,1]}^{n \times n}} \mathbb{E}_{\mathbf{A} \leftarrow N_\sigma(\bar{\mathbf{A}}), \mathbf{B} \leftarrow N_\sigma(\bar{\mathbf{B}})} [T_J(\mathbf{A}, \mathbf{B})],$$

where we use $\mathbf{A} \leftarrow N_\sigma(\bar{\mathbf{A}})$ to denote that $\mathbf{A}$ is a perturbation of $\bar{\mathbf{A}}$ according to $N_\sigma(\bar{\mathbf{A}})$.

---
[8]For the simplicity of presentation, in this section, we model the entries of payoff matrices and perturbations by real numbers. Of course, to connect with the complexity result of the previous section, where entries of matrices are in finite binary representations, we are mindful that some readers may prefer that we state our result and write the proof more explicitly using the finite binary representation. Using Equations (13) and (14) in the proof of Lemma 15, we can define a discrete version of the uniform and Gaussian perturbations and state and prove the same result.

An algorithm $J$ has a *polynomial smoothed time complexity* if for all $0 < \sigma < 1$ and for all positive integer $n$, there exist positive constants $c$, $k_1$ and $k_2$ such that

$$\text{Smoothed}_J[n, \sigma] \leq c \cdot n^{k_1} \sigma^{-k_2}.$$

The bimatrix game is in *smoothed polynomial time* if there exists an algorithm $J$ with polynomial smoothed time-complexity for computing a Nash equilibrium.

## 5.2 The Smoothed 2-NASH Conjecture

The following optimistic conjecture concerning the smoothed complexity of the Lemke-Howson algorithm for bimatrix games, made in [22], captures a repeatedly asked question after Spielman and Teng [23] proved that the smoothed complexity of the simplex method with the shadow-vertex pivoting rule is polynomial for solving linear programming.

**Conjecture 1 (Smoothed 2-NASH Conjecture).** *The problem of finding a Nash equilibrium of a bimatrix game can be solved in smoothed time polynomial in $n$ and $1/\sigma$, under uniform perturbations and Gaussian perturbations with magnitude $\sigma$ for all $0 < \sigma < 1$.*

Below, we show it is unlikely that this conjecture is true. We use the observation that the computation of a Nash equilibrium of a bimatrix game in the smoothed model can be used as a probabilistic polynomial reduction from the approximation problem of the bimatrix game to the search problem over a perturbed instance. Moreover, the approximation ratio is linearly related with the magnitude of the perturbation. Thus, if the smoothed complexity of an algorithm for the computation of a Nash equilibrium of bimatrix games is polynomial, then this algorithm can be used, with the help of perturbations, as a randomized polynomial-time algorithm for the approximation of Nash equilibrium [9]. So, we derive a hardness result of the bimatrix games in the smoothed model from our hardness result for the approximation of Nash equilibria.

## 5.3 Perturbation and Probabilistic Approximation

To help explain the probabilistic reduction from the approximation of bimatrix games to the solution of perturbed bimatrix games, we first define a notion of many-way polynomial reduction among **TFNP** problems.

**Definition 12 (Many-way Reduction).** *Let $\mathcal{F}$ be a set of polynomial-time computable functions and $g$ be a polynomial-time computable function. A search problem $Q_{R_1} \in$ **TFNP** is (F,g)-reducible to $Q_{R_2} \in$ **TFNP** if, for all $y \in \{0,1\}^*$, $(f(x), y) \in R_2$ implies $(x, g(y)) \in R_1$ for every input $x$ of $R_1$ and for every function $f \in \mathcal{F}$.*

We now show that if the smoothed complexity under uniform perturbations of the bimatrix game is low, then one can quickly find an approximate Nash equilibrium.

---

[9]This connection was previously discussed in [2] and [22].

**Lemma 15 (Smoothed Nash and Approximate Nash).** *If the problem of computing a Nash equilibrium of a bimatrix game is in smoothed polynomial time under uniform perturbations, then for any $0 < \epsilon < 1$, there exists a randomized algorithm for computing an $\epsilon$-approximate Nash equilibrium with expected time polynomial in $n$ and $1/\epsilon$.*

*Proof.* Suppose $J$ is an algorithm with polynomial smoothed complexity for computing a Nash equilibrium of a bimatrix game. Let $T_J(\mathbf{A}, \mathbf{B})$ be the complexity of algorithm $J$ for solving the bimatrix game defined by $(\mathbf{A}, \mathbf{B})$. Let $N_\sigma()$ denotes the uniform perturbation with magnitude $\sigma$. Then there exists constants $c$, $k_1$ and $k_2$ such that for all $0 < \sigma < 1$,

$$\max_{\bar{\mathbf{A}}, \bar{\mathbf{B}} \in \mathbb{R}^{n \times n}_{[-1,1]}} \mathrm{E}_{\mathbf{A} \leftarrow N_\sigma(\bar{\mathbf{A}}), \mathbf{B} \leftarrow N_\sigma(\bar{\mathbf{B}})} [T_J(\mathbf{A}, \mathbf{B})] \leq c \cdot n^{k_1} \sigma^{-k_2}.$$

For each pair of perturbation matrices $\mathbf{S}, \mathbf{T} \in \mathbb{R}^{n \times n}_{[-\sigma, \sigma]}$, we can define a function $f_{(\mathbf{S}, \mathbf{T})} : \mathbb{R}^{n \times n} \times \mathbb{R}^{n \times n} \to \mathbb{R}^{n \times n} \times \mathbb{R}^{n \times n}$ as $f_{(\mathbf{S}, \mathbf{T})}((\bar{\mathbf{A}}, \bar{\mathbf{B}})) = (\bar{\mathbf{A}} + \mathbf{S}, \bar{\mathbf{B}} + \mathbf{T})$. Let $\mathcal{F}_\sigma$ be the set of all such functions, i.e.,

$$\mathcal{F}_\sigma = \left\{ f_{(\mathbf{S}, \mathbf{T})} | \mathbf{S}, \mathbf{T} \in \mathbb{R}^{n \times n}_{[-\sigma, \sigma]} \right\}.$$

Let $g$ be the identity function from $\mathbb{R}^n \times \mathbb{R}^n$ to $\mathbb{R}^n \times \mathbb{R}^n$.

We now show that the problem of computing an $\epsilon$-approximate Nash equilibrium is $(\mathcal{F}_{\epsilon/2}, g)$-reducible to the problem of finding a Nash equilibrium of perturbed instances. More specifically, we prove that for every bimatrix game $(\bar{\mathbf{A}}, \bar{\mathbf{B}})$ and for every $f_{(\mathbf{S}, \mathbf{T})} \in \mathcal{F}_{\epsilon/2}$, an Nash equilibrium $(\mathbf{x}, \mathbf{y})$ of $f_{(\mathbf{S}, \mathbf{T})}((\bar{\mathbf{A}}, \bar{\mathbf{B}}))$ is an $\epsilon$-approximate Nash equilibrium of $(\bar{\mathbf{A}}, \bar{\mathbf{B}})$.

Let $\mathbf{A} = \bar{\mathbf{A}} + \mathbf{S}$ and $\mathbf{B} = \bar{\mathbf{B}} + \mathbf{T}$. Then,

$$|\mathbf{x}^T \mathbf{A} \mathbf{y} - \mathbf{x}^T \bar{\mathbf{A}} \mathbf{y}| = |\mathbf{x}^T \mathbf{S} \mathbf{y}| \leq \epsilon/2 \tag{13}$$

$$|\mathbf{x}^T \mathbf{B} \mathbf{y} - \mathbf{x}^T \bar{\mathbf{B}} \mathbf{y}| = |\mathbf{x}^T \mathbf{T} \mathbf{y}| \leq \epsilon/2. \tag{14}$$

Thus, for each Nash equilibrium $(\mathbf{x}, \mathbf{y})$ of $(\mathbf{A}, \mathbf{B})$, for any $(\mathbf{x}', \mathbf{y}')$,

$$(\mathbf{x}')^T \bar{\mathbf{A}} \mathbf{y} - \mathbf{x}^T \bar{\mathbf{A}} \mathbf{y} \leq \left( (\mathbf{x}')^T \mathbf{A} \mathbf{y} - \mathbf{x}^T \mathbf{A} \mathbf{y} \right) + \epsilon \leq \epsilon.$$

Similarly, $\mathbf{x}^T \bar{\mathbf{B}} \mathbf{y}' - \mathbf{x}^T \bar{\mathbf{B}} \mathbf{y} \leq \epsilon$. Therefore, $(\mathbf{x}, \mathbf{y})$ is an $\epsilon$-Nash equilibrium of game $(\bar{\mathbf{A}}, \bar{\mathbf{B}})$.

Now given the algorithm $J$ with polynomial smoothed time-complexity, we can apply the following randomized algorithm with the help of $(\mathcal{F}_{\epsilon/2}, g)$-many-way reduction to find an $\epsilon$-approximate Nash equilibrium of game $(\bar{\mathbf{A}}, \bar{\mathbf{B}})$:

**Algorithm** `smoothedForApproximation`$(\bar{\mathbf{A}}, \bar{\mathbf{B}})$

1. Randomly choose a pair of perturbation matrices $\mathbf{S}, \mathbf{T}$ of magnitude $\sigma$ and set $\mathbf{A} = \bar{\mathbf{A}} + \mathbf{S}$ and $\mathbf{B} = \bar{\mathbf{B}} + \mathbf{T}$.

2. Apply algorithm $J$ to find a Nash equilibrium $(\mathbf{x}, \mathbf{y})$ of $(\mathbf{A}, \mathbf{B})$.

3. Return $(\mathbf{x}, \mathbf{y})$.

The expected time complexity of `smoothedForApproximation` is bounded from above by the smoothed complexity of $J$ when the magnitude perturbations is $\epsilon/2$ and hence is at most $2^{k_2} c \cdot n^{k_1} \epsilon^{-k_2}$. □

We can similarly prove (with a slightly more complex argument to handle the low probability case when the absolute value of the perturbation is too large).

**Lemma 16 (Smoothed 2-NASH: Gaussian Perturbations).** *If the problem of computing a Nash equilibrium of a bimatrix game is in smoothed polynomial time under Gaussian perturbations, then for any $0 < \epsilon < 1$, there exists a randomized algorithm for computing an $\epsilon$-approximate Nash equilibrium with expected time polynomial in $n$ and $1/\epsilon$.*

**Remark 1.** *We observe that our proof of Lemma 15 can be extended to prove the following proposition.*

**Proposition 10 (Probabilistic Polynomial Reduction).** *Let $\mathcal{F}$ is a set of polynomial-time computable functions and $g$ be a polynomial-time computable function such that a* **TFNP** *search problem $Q_{R_1}$ is (F,g)-reducible to $Q_{R_2}$. For any positive function $t(n)$, if there exists a polynomial-time computable distribution $\mathcal{D}$ of $\mathcal{F}$ and an algorithm with time-complexity $T_{R_2}(y)$ for solving an instance $y$ of $R_2$ such that*

$$\max_{x \in I_n(R_1)} \mathrm{E}_{f \in_\mathcal{D} \mathcal{F}} [T_{R_2}(f(x))] \leq t(n),$$

*then, there there is a randomized algorithm for $R_1$ with expected complexity $t(n) + q(n)$, where $q(n)$ is the time for generating a function from $\mathcal{F}$ according to distribution $\mathcal{D}$.*

*In the statement above, $I_n(R_1)$ denotes the set of all inputs of $R_1$ of size $n$ and we use $f \in_\mathcal{D} \mathcal{F}$ to denote that $f$ is chosen from $\mathcal{F}$ according to distribution $\mathcal{D}$.*

## 5.4 Complexity Consequences

Setting $\epsilon = n^{-\Theta(1)}$, Theorem 3 and Lemmas 15 and 16 imply that the Smoothed 2-NASH Conjecture is unlikely to be true. Otherwise, we would have shown **PPAD** $\subset$ **RP**.

**Theorem 4 (Hardness of Smoothed Bimatrix Games).** *It is unlikely that the problem of computing a Nash equilibrium of a bimatrix game is in smoothed polynomial time, under uniform or Gaussian perturbations, unless* **PPAD** $\subset$ **RP**.

Thus, it is unlikely that the expected polynomial-time result of Barany, Vempala, and Vetta [2] can be extended to the smoothed model. In particular, our results show that the smoothed complexity of the Lemke-Howson algorithm is unlikely to be polynomial.

**Theorem 5 (Smoothed Complexity of Lemke-Howson).** *It is unlikely that the Lemke-Howson algorithm has a polynomial smoothed complexity (in $n$ and $1/\sigma$) under $\sigma$-uniform or $\sigma$-Gaussian perturbations, unless* **PPAD** $\subset$ **RP**.

# 6 Extensions and Conclusion

As the fixed-points and Nash equilibria are fundamental to many other search and optimization problems, our results and techniques may have a broader scope of applications and implications. Our hardness results can be naturally extended to $r$-person matrix games [19] and $r$-graphical games [11] for any fixed $r$.

## 6.1 Remarks

Although, we prove in this paper that it is **PPAD**-complete to compute a $1/n^6$-approximate Nash equilibrium of a two-person game, the constant 6 in our hardness result is not critical. In fact, it can be replaced by any constant $\delta > 0$. In other words, we can show

**Corollary 1.** *For any constant $\delta > 0$, the problem of computing a $1/n^\delta$-approximate Nash equilibrium of a normalized $n \times n$ bimatrix game is **PPAD**-complete.*

## 6.2 Open Questions

There remains a complexity gap on the approximation of Nash equilibria of bimatrix games: Lipton, Markakis, and Mehta [15] show that an $\epsilon$-approximate Nash equilibrium can be found in $n^{O(\log n/\epsilon^2)}$-time, while this paper shows that finding an $O(1/n^{\Theta(1)})$-approximate Nash equilibrium is **PPAD**-complete — hence it is unlikely that the bimatrix game has a fully polynomial-time approximation scheme. However, our hardness result does not cover the case when $\epsilon$ is an absolute constant between 0 and 1. Naturally, it is unlikely that finding an $\epsilon$-approximate Nash equilibrium of a bimatrix game is **PPAD**-complete when $\epsilon$ is an absolute constant, for otherwise, all **PPAD** problems would be solved in $n^{O(\log n)}$-time, due to the result of Lipton, Markakis, and Mehta.

Thinking optimistically, we would like to see the following conjecture to be true.

**Conjecture 2 (PTAS Approximate NASH).** *There is an algorithm to find an $\epsilon$-approximate Nash equilibrium of an $n \times n$ bimatrix game in time $O(n^{k+\epsilon^{-c}})$ for some positive constants $c$ and $k$.*

In the similar spirit, we would like to see a weaker version of the Smoothed 2-NASH Conjecture to be true.

**Conjecture 3 (Smoothed 2-NASH: Constant Perturbations).** *There is an algorithm to find a Nash equilibrium of an $n \times n$ bimatrix game with smoothed time complexity $O(n^{k+\sigma^{-c}})$ under perturbations with magnitude $\sigma$, for some positive constants $c$ and $k$.*

## A  Proof of Lemma 1

We start with some notations.

Let $f$ be a well-behaved function and $(C, 0^n)$ be an input instance of BROUWER$^f$, then $C$ is a valid Brouwer-mapping circuit with parameter $d$ and $\mathbf{r}$, where $d = \lfloor n/m \rfloor$, $m = f(n)$ and $r_i = 2^m$, $\forall\, i : 1 \le i \le d$.

A simplicial decomposition $\mathcal{S}$ of set $A_{\mathbf{r}}^d \subset \mathbb{R}^d$ is a collection of simplices such that

1. $\overline{A_{\mathbf{r}}^d} = \bigcup_{S \in \mathcal{S}} S$, where we use $\overline{A_{\mathbf{r}}^d}$ to denote the convex hull of $A_{\mathbf{r}}^d$;

2. for every $S \in \mathcal{S}$, if $S'$ is a face of $S$, then $S' \in \mathcal{S}$;

3. for every $n$-simplex $S \in \mathcal{S}$, its vertex set is accommodated;

4. for all $S_1, S_2 \in \mathcal{S}$, if $S_1 \bigcap S_2 \ne \emptyset$, then $S_1 \bigcap S_2$ is a face of both $S_1$ and $S_2$.

We now describe a simplicial decomposition of set $A_{\mathbf{r}}^d$. We let

$$V = \left\{ (\mathbf{p}, \sigma) \,\middle|\, \mathbf{p} \in \mathbb{Z}^d,\, 0 \le p_i < r_i - 1,\, \forall\, i : 1 \le i \le d,\, \sigma \text{ is a permutation over } \{1, 2, ..., d\} \right\}.$$

Obviously, $|V| \le 2^{md} d! = 2^{O(n \log n)}$. For every pair $v = (\mathbf{p}, \sigma) \in V$, it defines $d+1$ points $\mathbf{p}^0, \mathbf{p}^1 ..., \mathbf{p}^d$ in $\mathbb{Z}^d$ where $\mathbf{p}^0 = \mathbf{p}$ and $\mathbf{p}^{i+1} = \mathbf{p}^i + \mathbf{e}_{\sigma(i+1)}$, $\forall\, i : 0 \le i \le d-1$. We use $S_v$ to denote the $d$-simplex that is the convex hull of these $d+1$ points. One can check that

$$\mathcal{S} = \left\{ S \,\middle|\, \exists\, v \in V,\, S \text{ is a face of } S_v \right\}$$

is a simplicial decomposition of set $A_{\mathbf{r}}^d$. $\mathcal{S}$ also has the following nice property:

> For every $d$-simplex $S_{v_1} \in \mathcal{S}$, if $F$ is one of its facets, which is not on the boundary of $A_{\mathbf{r}}^d$, then we can find $S_{v_2} \in \mathcal{S}$ such that $F = S_{v_1} \bigcap S_{v_2}$ in polynomial time.

To prove that BROUWER$^f \in$ **PPAD**, we will construct a directed graph $G = (V \cup \{v^*\}, E)$. Both the in-degree and out-degree of every vertex are at most 1, and $v^*$ is a source of $G$. We will prove that, for every directed leaf $v$ except $v^*$, the vertex set of $S_v$ is a panchromatic set of $C$.

Let $F$ be a facet of some $d$-simplex $S_v \in \mathcal{S}$. $F$ is said to be panchromatic if it contains all the colors except "red". Let $\mathbf{p}^i$ be the vertex of $F$ which has color $i$, $\forall\, i : 1 \le i \le d$. We can define the orientation of facet $F$, that is, the orientation of the ordered sequence $\{\mathbf{p}^i\}_{1 \le i \le d}$, in $S_v$ in the standard way, which is either clockwise or counter-clockwise. The orientation defined has the following properties:

1. If $F = S_{v_1} \bigcap S_{v_2}$, then the orientations of $F$ in $S_{v_1}$ and $S_{v_2}$ are opposite;

2. Assume $S_v \in \mathcal{S}$ has all the colors except "red", then it has exactly two panchromatic facets $F_1$ and $F_2$ which have opposite orientations in it.

Besides, if $S_v$ contains exactly one panchromatic facet, then its vertex set is a panchromatic set of $C$.

Since $C$ is a valid Brouwer-mapping circuit, one can prove that there exists exactly one panchromatic $(d-1)$-simplex $F$ on the boundary of $\overline{A_{\mathbf{r}}^d}$. Here $F$ is the convex hull of $\mathbf{p}^d, ... \mathbf{p}^2, \mathbf{p}^1$, where $\mathbf{p}^d = 0$ and $\mathbf{p}^{i-1} = \mathbf{p}^i + \mathbf{e}_i$, for all $i : 2 \leq i \leq d$. $F$ is a facet of $d$-simplex $S_{v'} \in \mathcal{S}$ where $v' = (\mathbf{0}, \sigma')$ and $\sigma'(i) = d + 1 - i, \forall i : 1 \leq i \leq d$. Moreover, the orientation of $F$ in $S_{v'}$ is counter-clockwise.

Now we can construct the edge set $E$ of $G$ as follows. First, $v^* v' \in E$; Second, if $v_1, v_2 \in V$ and $S_{v_1} \bigcap S_{v_2} = F$ is a panchromatic $(d-1)$-simplex which is clockwise in $S_{v_1}$ (and thus, counter-clockwise in $S_{v_2}$), then $v_1 v_2 \in E$.

On the one hand, every directed leaf $v$ except $v^*$ gives a panchromatic set of $C$, that is, the vertex set of $S_v$. On the other hand, starting from $C$, we can compute a Turing machine $M$ in polynomial time, which generates the directed graph $G$.

In conclusion, we have BROUWER$^f \in$ **PPAD**.